\documentclass{ieeeaccess}

\usepackage{cite}
\usepackage{amsmath,amssymb,amsfonts}
\usepackage{algorithmic}
\usepackage{graphicx}
\usepackage{enumitem}
\usepackage{textcomp}
\usepackage{multirow}
\usepackage{colortbl}
\usepackage[font={sf,footnotesize}, labelfont={bf,color=accessblue}]{caption}
\usepackage{subcaption}
\usepackage{algorithmic}
\usepackage{algorithm}
\usepackage{threeparttable}
\usepackage{url}
\usepackage[switch]{lineno}

\def\BibTeX{{\rm B\kern-.05em{\sc i\kern-.025em b}\kern-.08em
    T\kern-.1667em\lower.7ex\hbox{E}\kern-.125emX}}

\begin{document}
\history{Date of publication August 16, 2021, date of current version August 24, 2021.}
\doi{10.1109/ACCESS.2021.3104520}

\title{FeSHI: \underline{Fe}ature Map-based \underline{S}tealthy \underline{H}ardware \underline{I}ntrinsic Attack}

\author{
    \uppercase{Tolulope A. Odetola}\authorrefmark{1}, 
    \uppercase{Faiq~Khalid}\authorrefmark{2},~\IEEEmembership{Member, IEEE},
    \uppercase{Hawzhin Mohammed}\authorrefmark{1}, 
    \uppercase{Travis C. Sandefur}\authorrefmark{1}, 
    \uppercase{Syed Rafay Hasan}\authorrefmark{1},~\IEEEmembership{Member, IEEE}
    \vspace{10pt}}
    
    \address[1]{Department of Electrical and Computer Engineering, Tennessee Technological University, Cookeville, TN, 38505, USA}
    \address[2]{Department of Computer Engineering, Technische Universit{\"a}t Wien (TU Wien), Vienna, Austria
}

\markboth
{Tolulope \headeretal: FeSHI: \underline{Fe}ature Map-based \underline{S}tealthy \underline{H}ardware \underline{I}ntrinsic Attack}
{Tolulope \headeretal: FeSHI: \underline{Fe}ature Map-based \underline{S}tealthy \underline{H}ardware \underline{I}ntrinsic Attack}

\tfootnote{%
This work is partially supported by the Carnegie Classification Funding from College of Engineering, and by the Faculty Research Grant from the Office of Research, Tennessee Tech University.
}

\begin{abstract}
    Convolutional Neural Networks (CNN) have shown impressive performance in computer vision, natural language processing, and many other applications, but they exhibit high computations and substantial memory requirements. To address these limitations, especially in resource-constrained devices, the use of cloud computing for CNNs is becoming more popular. This comes with privacy and latency concerns that have motivated the designers to develop embedded hardware accelerators for CNNs. However, designing a specialized accelerator increases the time-to-market and cost of production. Therefore, to reduce the time-to-market and access to state-of-the-art techniques, CNN hardware mapping and deployment on embedded accelerators are often outsourced to untrusted third parties, which is going to be more prevalent in futuristic artificial intelligence of things (AIoT) systems. These AIoT systems anticipates horizontal collaboration among different resource constrained AIoT node devices, where CNN layers are partitioned and these devices collaboratively compute complex CNN tasks. This horizontal collaboration opens another attack surface to the CNN-based application, like inserting the hardware Trojans (HT) into the embedded accelerators designed for the CNN. Therefore, there is a dire need to explore this attack surface for designing the secure embedded hardware accelerators for CNNs. Towards this goal, in this paper, we exploited this attack surface to propose an HT-based attack called FeSHI. Since in horizontal collaboration of RC AIoT devices different sections of CNN architectures are outsourced to different untrusted third parties, the attacker may not know the input image, but it has access to the layer-by-layer output feature maps information for the assigned sections of the CNN architecture. This attack exploits the statistical distribution, i.e., Gaussian distribution, of the layer-by-layer feature maps of the CNN to design two triggers for stealthy HT with a very low probability of triggering. Also three different novel, stealthy and effective trigger designs are proposed. To illustrate the effectiveness of the proposed attack, we deployed the LeNet and LeNet-3D on PYNQ to classify the MNIST and CIFAR-10 datasets, respectively, and tested FeSHI. The experimental results show that FeSHI utilizes up to 2\% extra LUTs, and the overall resource overhead is less than 1\% compared to the original designs. It is also demonstrated on the PYNQ board that FeSHI triggers the attack vary randomly making it extremely difficult to detect.
\end{abstract}
\titlepgskip=-25pt
\maketitle
%
\section{Introduction}
\label{sec:introduction}
Due to design flexibility through partial reconfiguration~\cite{du2019reliability,ayachi2021optimizing,zhang2021fpga,mo2021fpga,iqbal2020design} and fast prototyping~\cite{xu2018fclnn}\cite{huang2021fpga}, small-scale Field Programmable Gate Arrays (FPGAs)~\cite{liu20191} are being used to deploy Convolutional Neural Networks (CNNs) on Resource-Constrained (RC) devices in Artificial Intelligence of Things (AIoT), especially for Edge Intelligence (EI) based applications~\cite{zhang2019openei}. However, it is very challenging to achieve efficient deployment of CNNs on small-scale FPGAs. Therefore, to achieve resource efficiency, short time-to-market, and state-of-the-art implementation, the deployment of pre-trained CNN on hardware accelerators can be outsourced to untrusted third parties, which provide soft IPs or hard IPs (in the form of bitstream files) of the deployed CNN. Although outsourcing the CNN deployment achieves the required goal, it comes with security risks such as malicious hardware insertions, which can be stealthy by nature~\cite{odetola2021sowaf}. 

\begin{table*}[!t]
    \centering
    \caption{A brief comparison of of the state-of-the-art HT Insertion techniques for CNN}
    \label{tab:comparison_table}
    \resizebox{1\linewidth}{!}{
        \begin{tabular}{|l|c|c|c|c|c|c|c|>{\columncolor[RGB]{200, 240, 255}}c|}
        \hline
        \textbf{Features / Assumptions} & \cite{liu2017trojaning} & \cite{clements2018hardware} & \cite{clements2019hardware} & \cite{zhao2019memory} & \cite{liu2020sequence} & \cite{hailesellasie2019vaws}  & \cite{zou2018potrojan} & \textbf{FeSHI}\\ \hline
        Access to full CNN architecture to evaluate the triggering probability & \checkmark & \checkmark & \checkmark & \checkmark & \checkmark & \checkmark & \checkmark & $\times$\\ \hline
        Pre-defined sequence for designing a trigger & $\times$ & $\times$ & $\times$ & $\times$ & \checkmark & $\times$ & $\times$ & $\times$ \\ \hline
        Weight manipulation & $\times$ & $\times$ & $\times$ & \checkmark & $\times$ & \checkmark & $\times$  & $\times$ \\ \hline
        Requires knowledge of the input Image & $\times$ & $\times$ & $\times$ &  \checkmark & $\times$ & $\times$ & $\times$  & $\times$\\ \hline
        Requires access to internal of mathematical operations & $\times$ & \checkmark & $\times$ & $\times$ & $\times$ &$\times$ & $\times$ & $\times$\\ \hline
        Number of FPGAs / resource-constrained IoT devices for the CNN Inference & one & one & one & one & one & one & one & one/multiple \\ \hline
        \end{tabular}
    }
\end{table*}

Recently, various hardware Trojans (HTs) insertion techniques in CNNs have been explored, but most of them are applicable under the following assumptions:

\begin{itemize}[leftmargin=*]
    \item These techniques require full access to the CNN architecture to evaluate triggering probability, stealthiness, and resource utilization (comparing them with design constraints of their attacks~\cite{liu2017trojaning,clements2018hardware,clements2019hardware}).
    \item Few of the techniques require a pre-defined sequence, where a sequence of events serves as the trigger for the attack~\cite{liu2020sequence}. This sequence of events also requires knowledge of the full CNN architecture, and in the absence of information about any layer, especially the initial layer and/or the last layer, there is a high probability that the sequence of events may never occur.
    \item Some of the techniques require weight manipulation, which requires a change in the values of the weights of the CNN to achieve misclassification that can easily be detected using model integrity analysis~\cite{zhao2019memory,hailesellasie2019vaws}.
    \item Another work requires knowledge of input dataset or access to the input image to introduce the trigger~\cite{zhao2019memory}.
    \item In \cite{clements2018hardware}, Clements et al. modify the internal structure of certain operations to inject malicious behavior, which means they require that the attacker has access to internal mathematical operations. 
\end{itemize}

In the real-world scenario, it is impossible to get full access to the CNN architecture~\cite{marchisio2019deep,zhang2019building,shafique2018overview,hanif2018robust,kriebel2018robustness,khalid2019trisec,khalid2019red,khalid2019fademl}. Therefore, most of these HT attacks are applicable under the limited attack scenario. Moreover, most of the approaches adopted in state-of-the-art hardware/firmware Trojan attacks on hardware accelerator-based CNN inference are focused on the deployment of CNN on a single FPGA with access to the complete CNN pipeline~\cite{clements2018hardware,zou2018potrojan,liu2020sequence}. Table~\ref{tab:comparison_table} summarizes the differences in the state-of-the-art HT insertion techniques.

\begin{figure}[!t]
	\centering
	\includegraphics[width=1\linewidth]{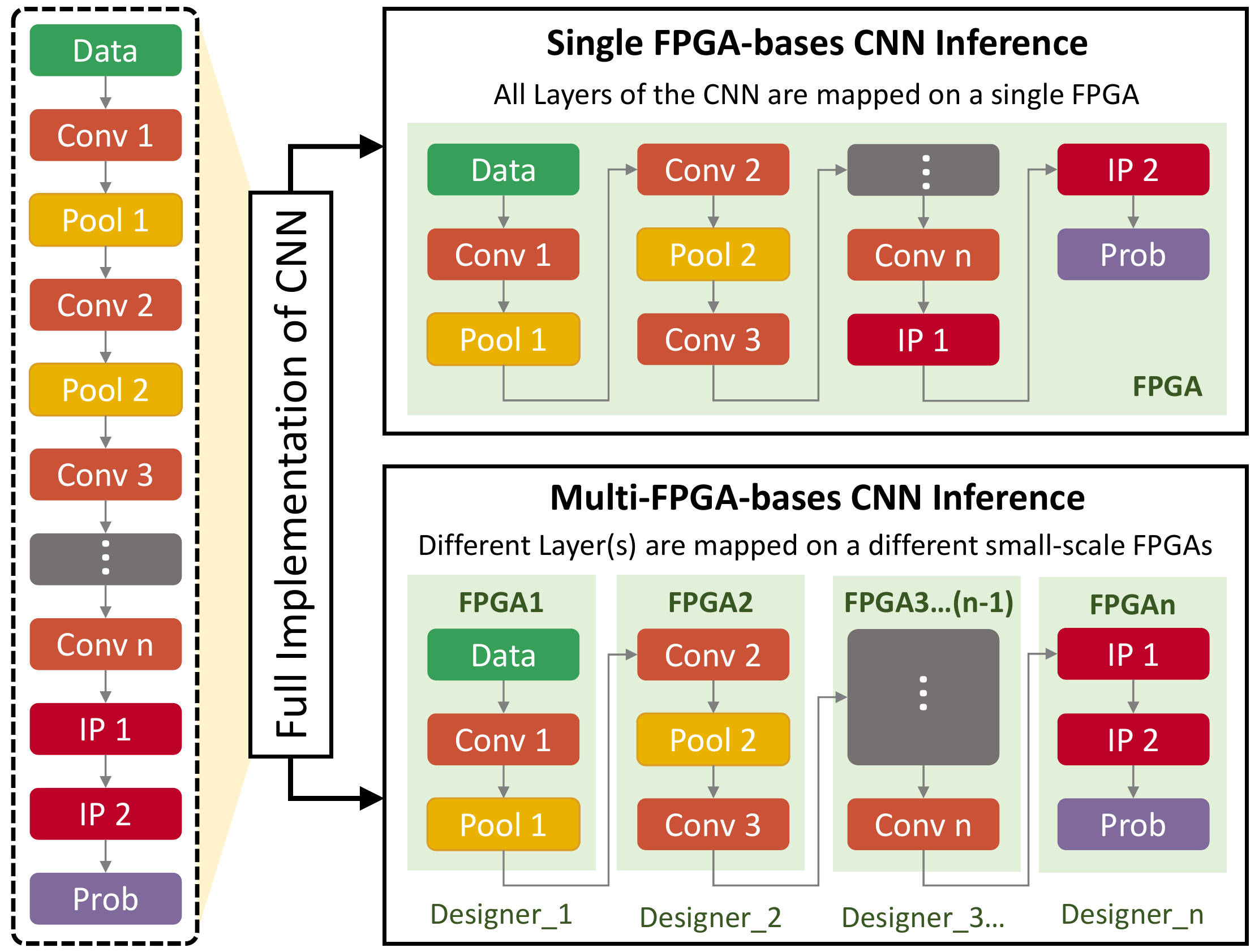} 
	\caption{A figurative comparison between single FPGA-based CNN inference and multi-FPGA-based CNN. In single FPGA-based inference, all CNN layers are implemented on an FPGA, and the attacker has access to the complete CNN architecture. On the other hand, in multi-FPGA-based inference, full CNN is divided into multiple different partitions, where different third-party IP designers can design each partition. These partitions can be mapped on multiple FPGAs. }
	\label{fig:attack_scenario}
\end{figure}

\begin{figure}[!t]
	\centering
	\includegraphics[width=1\linewidth]{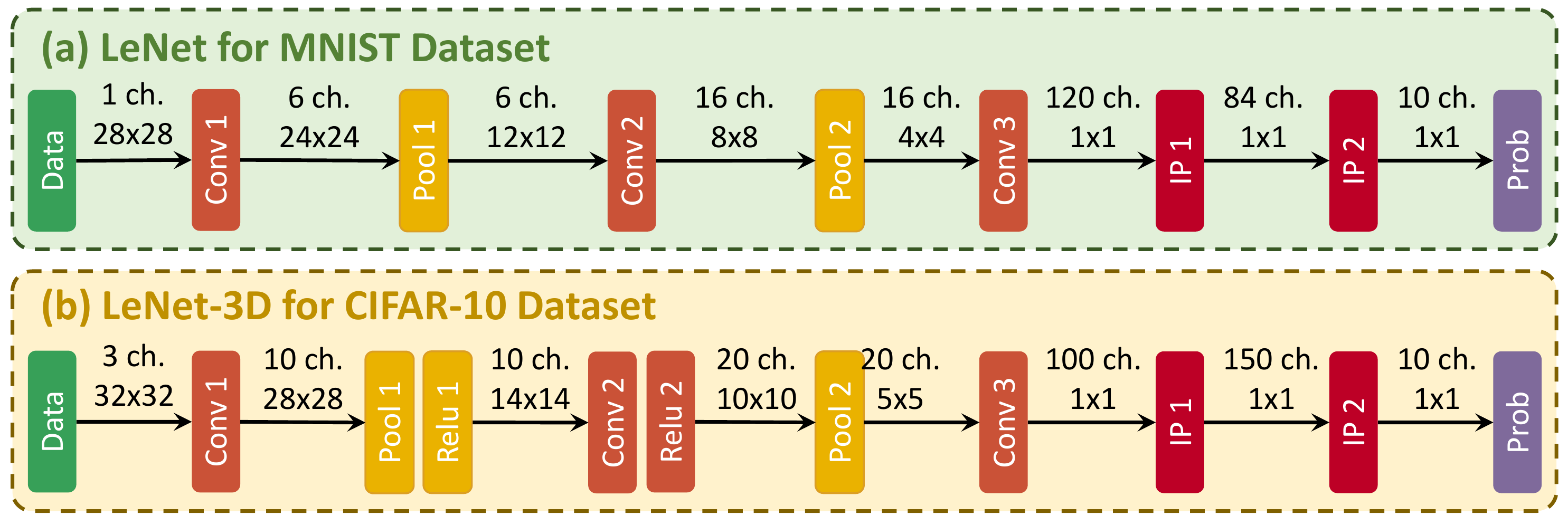} 
	\caption{CNN models used in experiments; (a) LeNet trained/ tested on MNIST (b) LeNet-3D trained/ tested on CIFAR-10.}
	\label{fig:CNN_MOdel}
\end{figure}

\begin{figure*}[!t]
	\centering
	\includegraphics[width=1\linewidth]{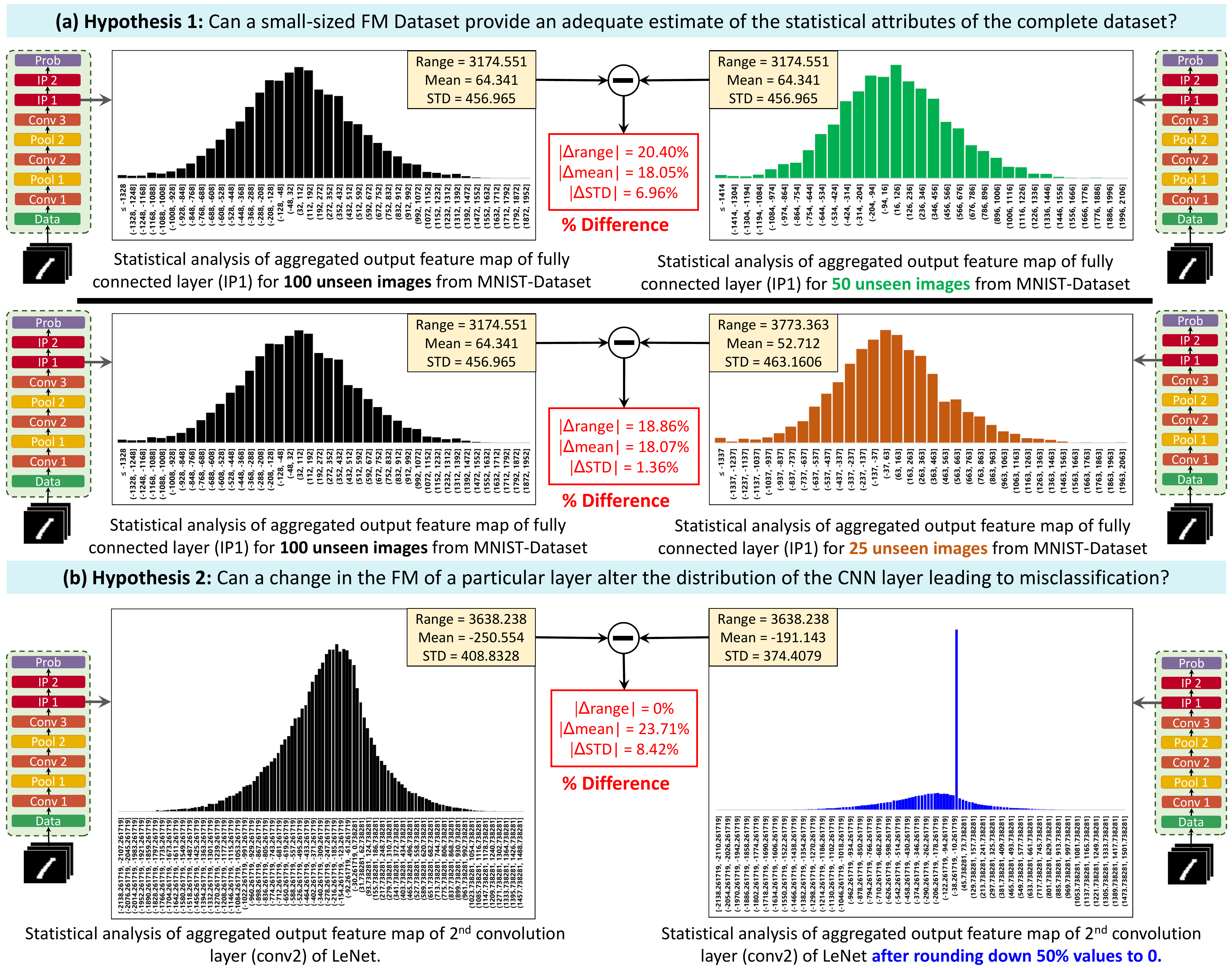} 
	\caption{(a) \textbf{Hypothesis 1 (Feature map analysis)}: Gaussian distribution plots of aggregated output feature map of the fully connected layer ($ip1$ layer of LeNet CNN model shown in Fig.~\ref{fig:CNN_MOdel}) in response to 100 and 50 MNIST images, respectively. The plots depict that the Gaussian plot of the response to 50 MNIST images similarity to the plot in response to 100 MNIST images. The plots show an approximate estimate in terms of Range, Mean and Standard deviations. (b) \textbf{Hypothesis 2:} Change in the Gaussian distribution of the output feature map of the second convolution layer of LeNet ($conv2$ layer of LeNet CNN model shown in Fig.~\ref{fig:CNN_MOdel}) as a result of rounding down $50\%$ of the values of the feature maps. \textit{The plot shows similar statistical attributes in range, mean and standard deviation despite drastic changes in the Gaussian distribution plot.}}
	\label{fig:MOtivational_Analysis}
\end{figure*}

To address some of these attacks, defense techniques such as \cite{zhang2021tad} is proposed, which is a trigger approximation-based Trojan detection framework. It searches for triggers in input space and also detects Trojans in a feature map space. Gao et al. \cite{gao2019strip} propose a STRIP defense methodology that turns the strength of an input-agnostic trigger into a means of detecting Trojans at run-time. Doan et al. \cite{doan2020februus} propose Februus that neutralizes Trojan attacks on Deep Neural Network (DNN) systems at run-time. 

Partitioning of CNN with horizontal collaboration among RC AIoT device ~\cite{hadidi2018distributed,hu2020fast,mao2017modnn,mao2017mednn} or via vertical collaboration \cite{shen2017maximizing} \cite{zeng2019boomerang} is also considered as an alternative approach to defend attacks. This is because partitioned CNN acts like a distributed network; hence it contains the inherent security feature of such networks. Moreover, with partitioned CNN on multiple FPGAs, third parties (3Ps) do not have access to the full CNN pipeline rendering current attack techniques invalid. However, partitioning CNNs on multiple FPGAs can also pose security concerns.

It is possible that one or more of these FPGA devices are malicious, and the CNN deployment on FPGA may be compromised. However, since malicious devices do not have access to the complete network or ground truth, coming up with stealthy and effective hardware attacks is challenging. In this paper, we investigate and demonstrate that an attacker may carry out an attack on the CNN while not having complete knowledge of CNN architecture and training or testing dataset. It is worth mentioning here that though the attacker does not have access to the training dataset, it may still be provided with a Feature Map-based Validation Dataset (FMV-Dataset) to verify the functional correctness hardware. Please note, it is not the traditional validation dataset as provided along with the training dataset for any supervised machine learning techniques.

\subsection{Motivational Case Study}

To investigate the viability of the above-mentioned attack, we need to explore whether the small-sized FMV-Dataset, with a possible payload in only one of the partitioned CNN layers, can lead to an effective and stealthy attack or not. To answer this, we postulate two hypotheses and performed two experiments to confirm the validity of our hypotheses: 

\textcolor{accessblue}{\textbf{Hypothesis 1:} A small-sized FMV-Dataset can provide an adequate estimate of the statistical attributes of the complete dataset}

To investigate this hypothesis, we take 100 images from the MNIST dataset (as our ground truth) as input into LeNet CNN shown in Fig.~\ref{fig:MOtivational_Analysis}. The feature maps of the first fully connected layer (ip1 in Fig.~\ref{fig:MOtivational_Analysis}) are aggregated, and the Gaussian distribution plotted. This is repeated for 50 images from the MNIST dataset and compared with the ground truth. From Fig.~\ref{fig:MOtivational_Analysis}, we observed that the behavior learned from the larger dataset is very similar to the behavior learned using a smaller dataset. For example, Fig.~\ref{fig:MOtivational_Analysis} shows the histogram plot of aggregated result of convolution of $(100 \times 84 \times 1)$ input vectors against $(50 \times 84 \times 1)$. The plots show similar Gaussian distribution and statistical attributes in terms of range, mean and standard deviation with a $50\%$ reduction in the dataset (from 100 images to 50 images). These histograms show that the range of values (RoVs) on the $x-axis$ remains very similar for the two cases (only 2.73\%). Hence with this result, we validate our hypothesis. In the actual trigger design, we propose to exploit this similarity of RoVs.

\textcolor{accessblue}{\textbf{Hypothesis 2:} A change in the feature map of a particular layer can alter the distribution of the CNN layer leading to misclassification}

To demonstrate this, we experimented by rounding down a random amount of values across all the channels in the feature map of the layer of interest to zero (in this case, it is convolution layer). For example, Fig.~\ref{fig:MOtivational_Analysis} illustrates the histogram plot of aggregated result of the convolution of $(10\times 16 \times 12 \times 12)$ input vectors (10 indicates the number of feature maps) with $(16 \times 5 \times 5)$ weight matrices before and after the rounding down half of the elements of the resulting feature maps. The result shows similar statistical attributes with small differences in the range, mean and standard deviations despite the drastic change in the shape of the Gaussian distribution of the feature map that can serve as input to subsequent layers. Due to the change in Gaussian distribution, we conclude that it is worth exploring to see the effect of such an analysis on complete CNN. Hence, this toy example inspired us to proceed with this approach for the design of a stealthy payload.

\subsection{Research Challenges}
The design of an attack targeted at CNN inference on multiple FPGAs leads to the following research challenges:
\begin{enumerate}[leftmargin=*]
    \item How can we scientifically implement a stealthy attack trigger, based on the statistical attributes of a CNN layer, without access to the full CNN architecture?  
    \item How can we design a payload that will cause misclassification at the final output of CNN,  having access to only one or a few of the CNN partitioned layers in a horizontal collaboration-based AIoT system? 
\end{enumerate}

\subsection{Novel Contributions}

\begin{figure}[h]
    \centering
    \includegraphics[width=1\linewidth]{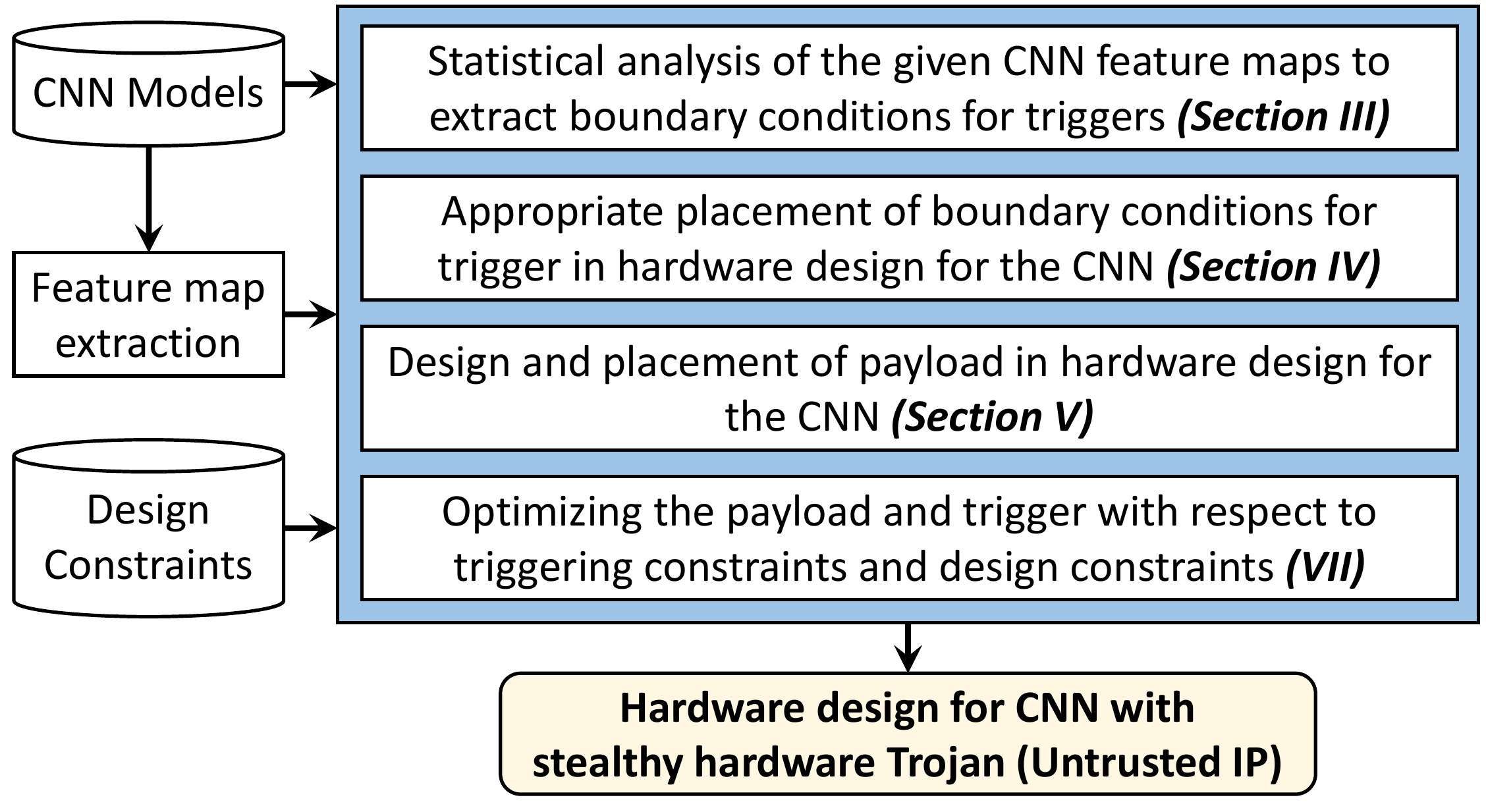}
    \caption{FeSHI attack methodology showing the steps in the deployment of CNN models that involves the output feature map extraction and analysis. The analysis shows the selection of triggering boundary conditions (TBCs) and insertion into the CNN hardware design, payload design, and optimization.}
    \label{novel_diagram}
\end{figure}

To address the above research challenges, we propose a novel Trojan attack methodology for FPGA-based CNN inference called FeSHI (Feature map Based Stealthy Hardware Intrinsic Attack). The FeSHI methodology is shown in Fig.~\ref{novel_diagram}. 
Our contributions include the following:
\begin{enumerate}[leftmargin=*]
    \item We propose an attack that triggers at rare, occurring events without requiring the complete knowledge of CNN architecture and with no access to weights (except for the partitioned layer(s) that attacker has been provided with). The triggers are designed based on the statistical exploitation of the CNN feature maps to identify vulnerable triggering boundary conditions (TBCs). In this work, TBCs are a range of values that lie beyond the $3\sigma$ values of a distribution. (See Section ~\ref{attack_meth} and~\ref{trigger}).  
    \item We also propose a dynamic payload mechanism to further improve the stealthiness of the attack (if the defender tries to come up with test-vectors to detect hardware attacks due to such rarely occurring events).
\end{enumerate}
The evaluation of our proposed FeSHI attack is carried out on the Xilinx PYNQ FPGA board (ZYNQ 7020). The experiment test bench shows the following:
\begin{itemize}[leftmargin=*]
    \item FeSHI proved to be very stealthy by requiring less than $1\%$ additional hardware resources with less than $1\%$ addition to the latency (See Section~\ref{results})
    \item Misclassification (associated payload) can be achieved (after activation) with the attack placed in any of the CNN layers despite limited information about the position of the layer in the CNN architecture on multiple FPGAs.
\end{itemize}

\section{Preliminaries}
To increase the readability and understanding of the paper, in this section, we provide the description of the assumed threat model along with all the assumptions, the nomenclature of DNN parameters, and the CNN models used in this paper.   
\subsection{Threat model}

\begin{figure}[h]
    \centering
    \includegraphics[width=1\linewidth]{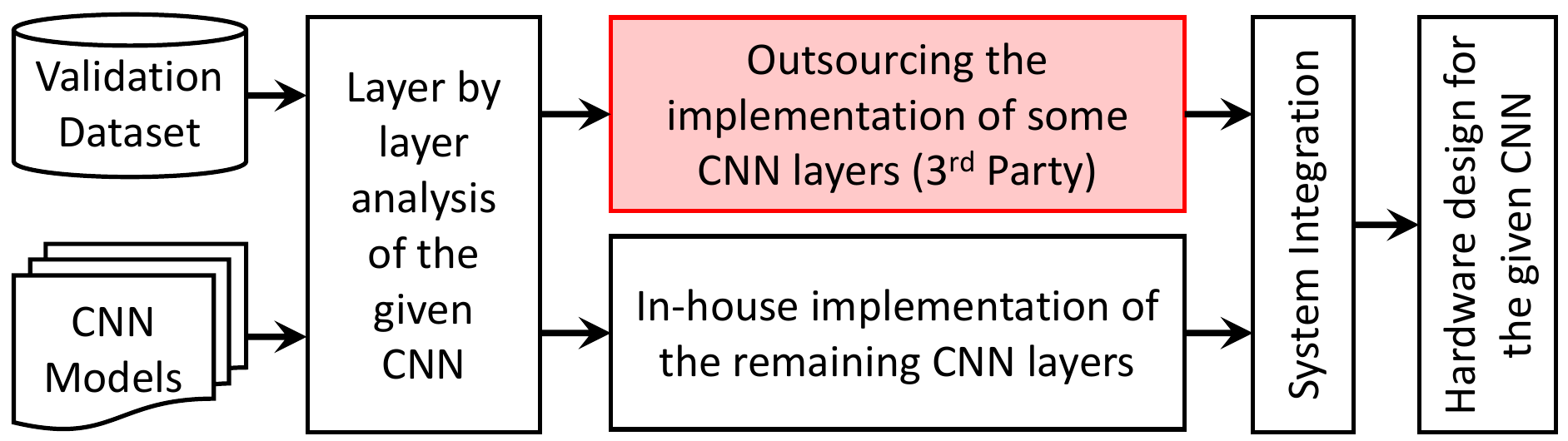}
    \caption{Overview of the design time mapping of CNN Model showing compromised outsourced layers by 3Ps bitstreams integrated trusted in-house implementation of trusted layers.}
    \label{threat_model}
\end{figure}

The threat model adopted in this paper follows a gray-box attack where the adversary only has limited knowledge about the CNN architecture. It is assumed that the attacker has no knowledge of the testing and validation data samples required for the supervised machine learning of CNN models. This threat model assumes that the CNN can be partitioned into different blocks by employing horizontal offloading, as alluded in Fig.~\ref{fig:attack_scenario} and outsourced to different third parties. In this attack, the attacker has no access to the full CNN layers (specifically the first and final layers, which are assumed to be implemented at a trusted IP integration facility - referred to as In-house implementation in Fig.~\ref{threat_model}), rather the attacker has access to only partial layers.

\begin{figure}[h]
    \centering
    \includegraphics[width=1\linewidth]{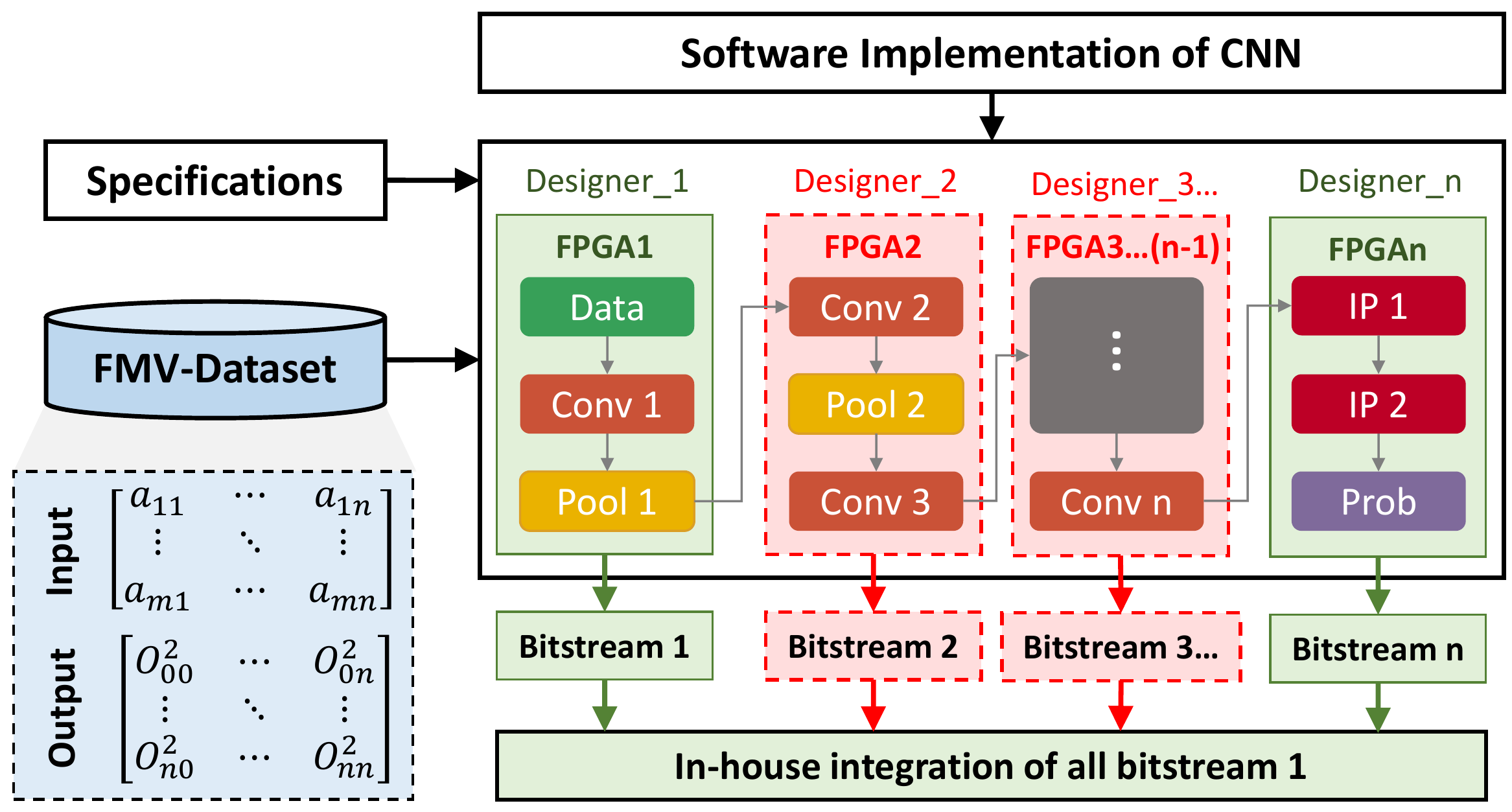}
    \caption{Overview of the CNN deployment methodology showing untrusted sections accessible to the third party.}
    \label{threat_model1b}
\end{figure}

\begin{figure}[h]
    \centering
    \includegraphics[width=1\linewidth]{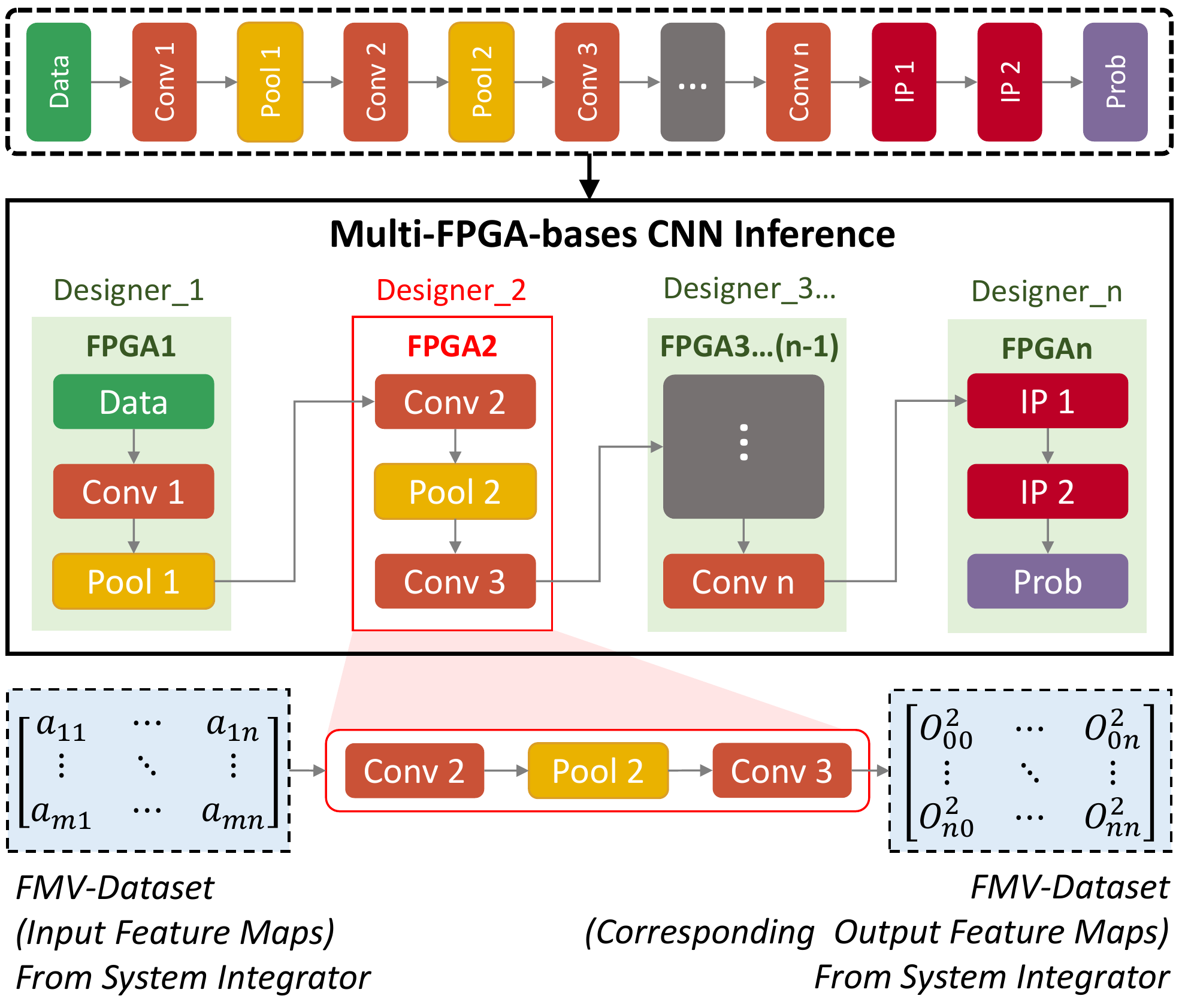}
    \caption{Threat Model shows the usage of FMV-Dataset for validation and access to the layer-by-layer output of the given architecture.}
    \label{threat_model1a}
\end{figure}

Fig.~\ref{threat_model1b} shows a CNN architecture implemented on $n$ number of FPGAs. The initial and final layers of the full CNN architecture are designed and implemented by trusted parties to generate trusted bitstream files. Middle layers of the CNN ($conv_2$ to $conv_n$) are outsourced to untrusted 3Ps in which any one of them may be an attacker, and hence their corresponding bitstream files are untrusted. From Fig.~\ref{threat_model1a}, FPGA $2$ is an attacker. Hence, all the trusted and untrusted bitstream files are integrated into the distributed CNN design as shown in Fig.~\ref{threat_model1b}. The attacker can perform layer-by-layer analysis on the outsourced partial layers they have access to. We assume that the 3P vendors are not trustworthy, as shown in Fig.~\ref{threat_model}, and 3PIPs may contain Hardware Trojan. The attacker provides the CNN hardware design as a bitstream file to the trustworthy project owner or system IP integrator (for single or multi-FPGA CNN deployment).

Fig.~\ref{threat_model} depicts the trusted and untrusted aspects of the CNN hardware design. The 3P designer has access to the untrusted sections in the IP generation, as shown in Fig.~\ref{threat_model1b}. This work also assumes that for functional/behavioral verification purposes, an FMV-Dataset is provided to the 3P designer \cite{odetola20192l}, without knowledge or access to the first or final layers. In this work, we make the assumption that the FMV-Dataset is a validation dataset that contains several sets of data instances made up of input feature maps (input to the partial CNN architecture provided to the untrusted 3P) and the corresponding output feature map (output of the entire partial architecture provided to the untrusted 3P) as shown in Fig.~\ref{threat_model1a}. During functional verification, using the input and output data instances from the FMV-Dataset, the 3P has access to the layer-by-layer output feature map, which serves as subsequent input layers. The untrusted 3P can perform statistical analysis on the layer-by-layer output of each respective layer to design a stealthy trigger. Hence, the FMV-Dataset provides a means of validating the design from the project owner`s (defender) perspective but also gives the untrusted 3P insights into the range of values of the intermediate layers to exploit the vulnerabilities in the CNN design.

\subsection{CNN Parameter Naming Convention}
\label{convention}
\begin{figure}[h]
    \centering
    \includegraphics[width=1\linewidth]{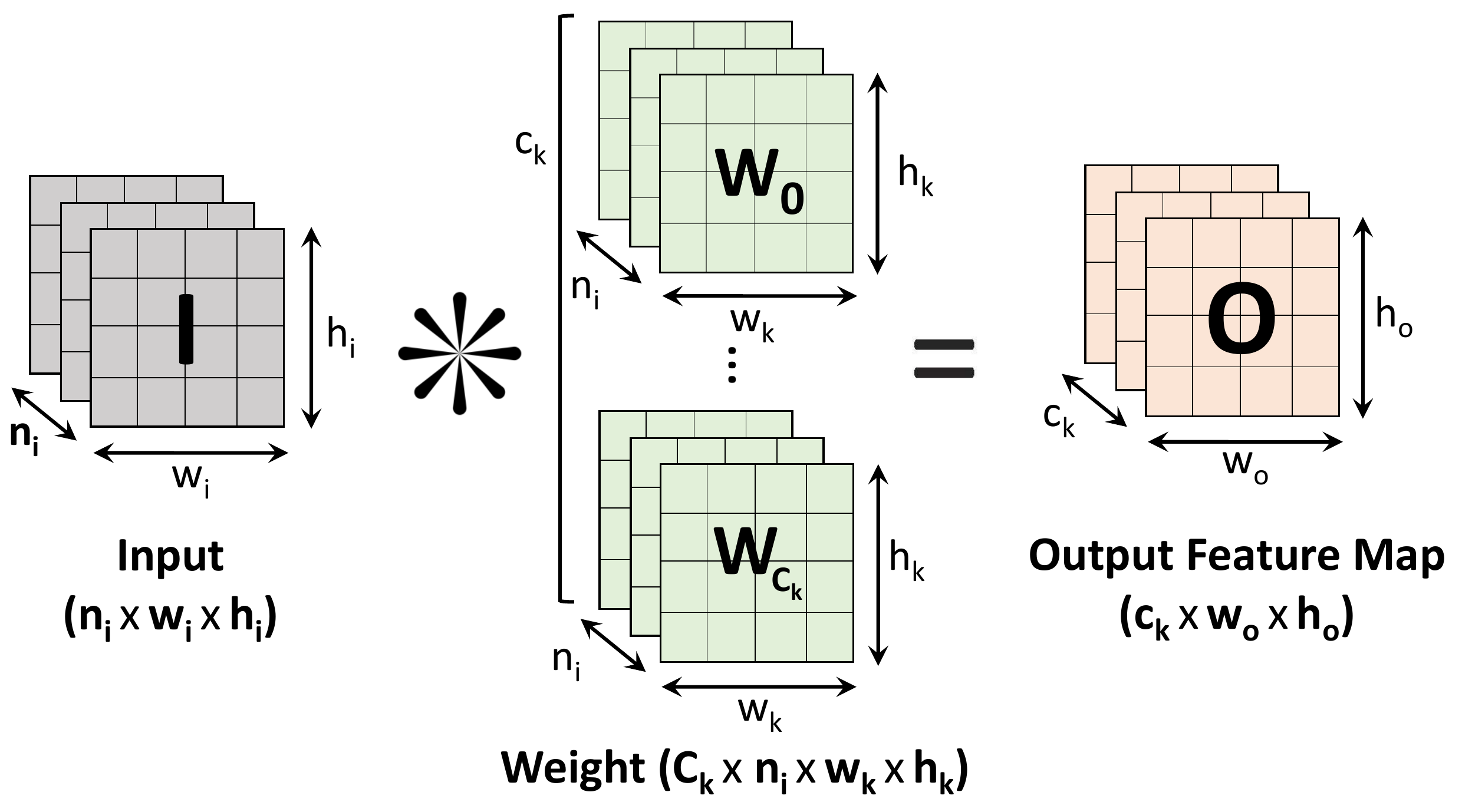}
    \caption{Conceptual representation for dimensions of convolution showing the naming conventions used in this paper}
    \label{dimension}
\end{figure}

Fig.~\ref{dimension} shows a depiction of 3D convolution between an input image $I$ and weights matrix ($W_0, ..., W_k$). In this paper, we provide several techniques pertaining to the attack trigger and payload. To understand this paper better, it is pertinent that we define our parameters.

\begin{itemize}
    \item $n_i$: Number of input image channels.
    \item $w_i$: Number of columns of the input image.
    \item $h_i$: Number of rows of the input image.
    \item $w_k$: Number of columns of the weight matrix.
    \item $h_k$: Number of rows of the weight matrix.
    \item $c_k$: Number of the weight matrix.
    \item $w_o$: Number of columns of the output feature map.
    \item $h_o$: Number of rows of the output feature map.
\end{itemize}

\subsection{CNN Models}
In this paper, we used some of the most commonly used CNN models. Fig.~\ref{fig:CNN_MOdel} shows these CNN models (LeNet and LeNet-3D) used for the classification of MNIST and Cifar10 datasets, respectively. LeNet CNN consist of $6$ layers from convolution layer ($conv1$) to fully connected layer ($ip1$) excluding the classification layer ($ip2$) used for the prediction of MNIST datasets. LeNet-3D CNN is similar to LeNet CNN for the classification of the Cifar-10 dataset, with the exception that the input image is RGB instead of grayscale.

\section{\underline{Fe}ature Map Based \underline{S}tealthy \underline{H}ardware \underline{I}ntrinsic (F\MakeLowercase{e}SHI) Attack}
\label{attack_meth}
FeSHI makes use of statistical attributes of the Gaussian distribution of the feature maps of CNN layers to investigate rarely occurring range of values (RoVs) for the design of a stealthy hardware intrinsic attack. This section provides the premise of the FeSHI attack's methodology and gives an overview of the attack methodology. 

\subsection{Methodology Premise}

In CNNs, convolution layers are the most computation-intensive while the fully connected layers are most memory demanding \cite{sun2017power}. CNN architectures are designed such that layer-by-layer outputs (feature maps) of previous layers serve as inputs to subsequent layers to achieve classification, as shown in Fig.~\ref{fig:attack_scenario}.

\subsubsection*{Feature Maps Follow Gaussian Distribution}
To achieve a successful implementation of this attack, without losing generality, we consider a convolution layer with the following mathematical operations:
\begin{itemize}
    \item let $I$ be an input vector of shape $n_i \times h_i \times w_i$
    \item let $W$ be a kernel matrix of shape $c_k \times n_k \times h_k \times w_k$
    \item let $F$ be the feature map of the convolution layer of shape $c_k \times w_o \times h_o$
\end{itemize}
Hence, the feature maps can be obtained by $F = I \circledast W $, where $\circledast$ represent convolution operation. Therefore, feature maps can be mathematically modeled as:
\begin{equation}
    (n_i \times h_i \times w_i) \circledast (c_k\times n_k \times h_k \times w_k) = (c_k \times w_o \times h_o)
\end{equation}

Many common CNN frameworks make use of Xavier (Glorot) \cite{glorot2010understanding} as the default weight initializers \cite{Jihongcaffe} for training CNN architectures \cite{Machinelearninginitializers}. The weights from Xavier initializers follow a Gaussian distribution \cite{yedroudj2018yedroudj}. Input images or feature maps also follow Gaussian distribution \cite{ArcHistogram}. Hence: 
\begin{equation}
     W \sim \mathcal{N}(\mu,\,\sigma^{2})
\end{equation}
\begin{equation}
     I \sim \mathcal{N}(\mu,\,\sigma^{2})
\end{equation}
where $\mu$ and $\sigma^{2}$ represent the mean and variance of the distributions respectively. Convolution operation between these weights ($W$) and input images or vectors ($I$) results in feature maps ($F$) that also follow Gaussian distribution according to Cramer's theorem \cite{Wiki_normal}. Therefore, based on corollary of Cramer's theorem feature can be represented as:
\begin{equation}
     I \circledast W = F \sim \mathcal{N}(\mu,\,\sigma^{2})
\label{eqn_core}
\end{equation}

\begin{figure}[!t]
    \begin{subfigure}{\linewidth}
      \centering
      \includegraphics[width=0.6\linewidth]{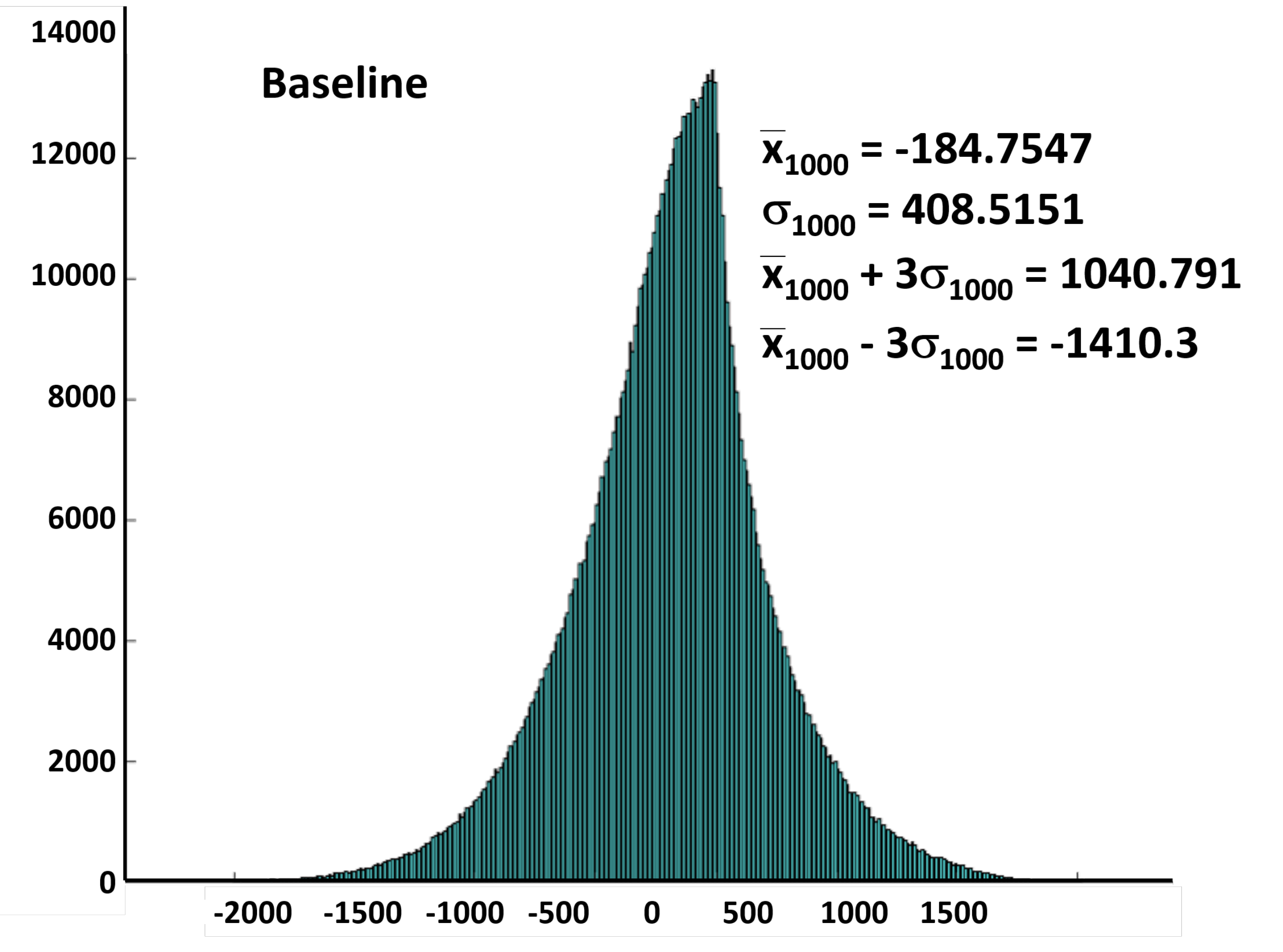}  
      \caption{Gaussian distribution, mean, standard deviation and $3\sigma$ values of $conv1$ output in response to 1000 input MNIST baseline images.}
      \label{input_vectors1_a}
    \end{subfigure}
    \begin{subfigure}{\linewidth}
      \centering
      \includegraphics[width=1\linewidth]{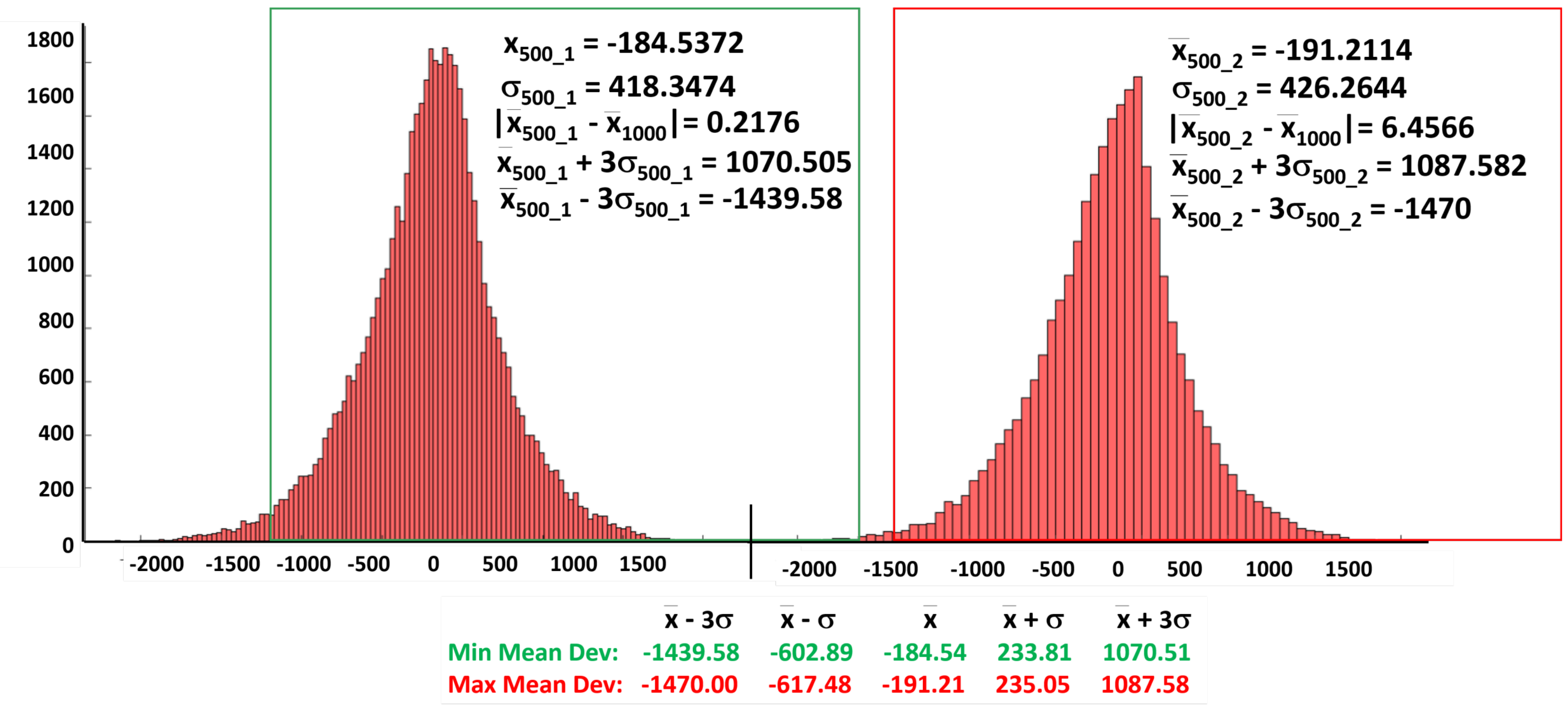}  
      \caption{Gaussian distribution, mean and standard deviation for 2x500 input vectors and showing plots with the maximum (in Red box) and minimum (in Green box) mean deviation of $3\sigma$ values of $conv1$ output for the 2x500 MNIST images when compared with baseline (the maximum $3\sigma$ deviation is within $4.50\%$ of the baseline values).}
      \label{input_vectors1_b}
    \end{subfigure}
    \begin{subfigure}{\linewidth}
      \centering
      \includegraphics[width=1\linewidth]{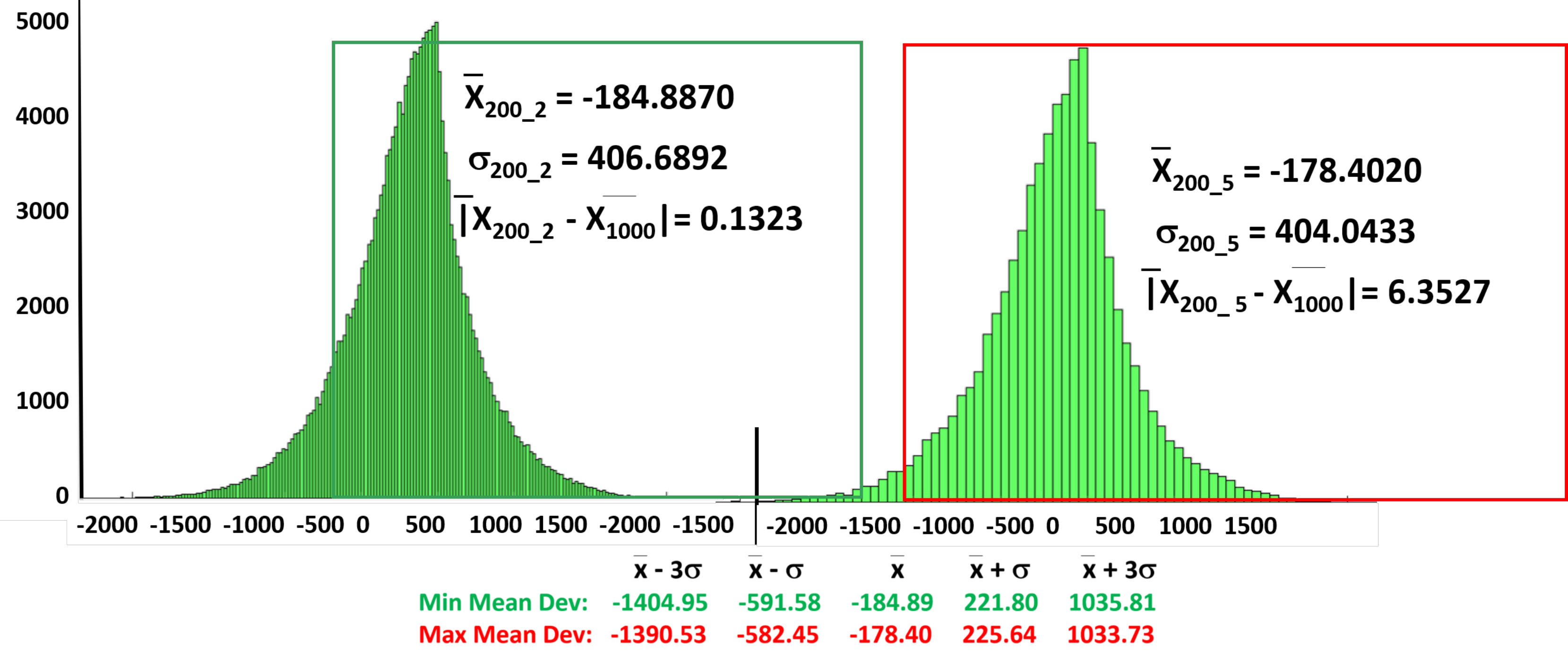}  
      \caption{Gaussian distribution, mean and standard deviation for 5x200 input vectors showing plots with the maximum (in Red box) and minimum (in Green box) mean deviation of $3\sigma$ values of $conv1$ output for the 5x200 MNIST images compared with baseline (the maximum $3\sigma$ deviation is within $1.40\%$ of the baseline values). The full figure showing the distribution and statistical properties of all the 5 buckets each of 200 images is represented in Appendix\ref{appendix}.}
      \label{input_vectors1_c}
    \end{subfigure}
    \begin{subfigure}{\linewidth}
      \centering
      \includegraphics[width=1\linewidth]{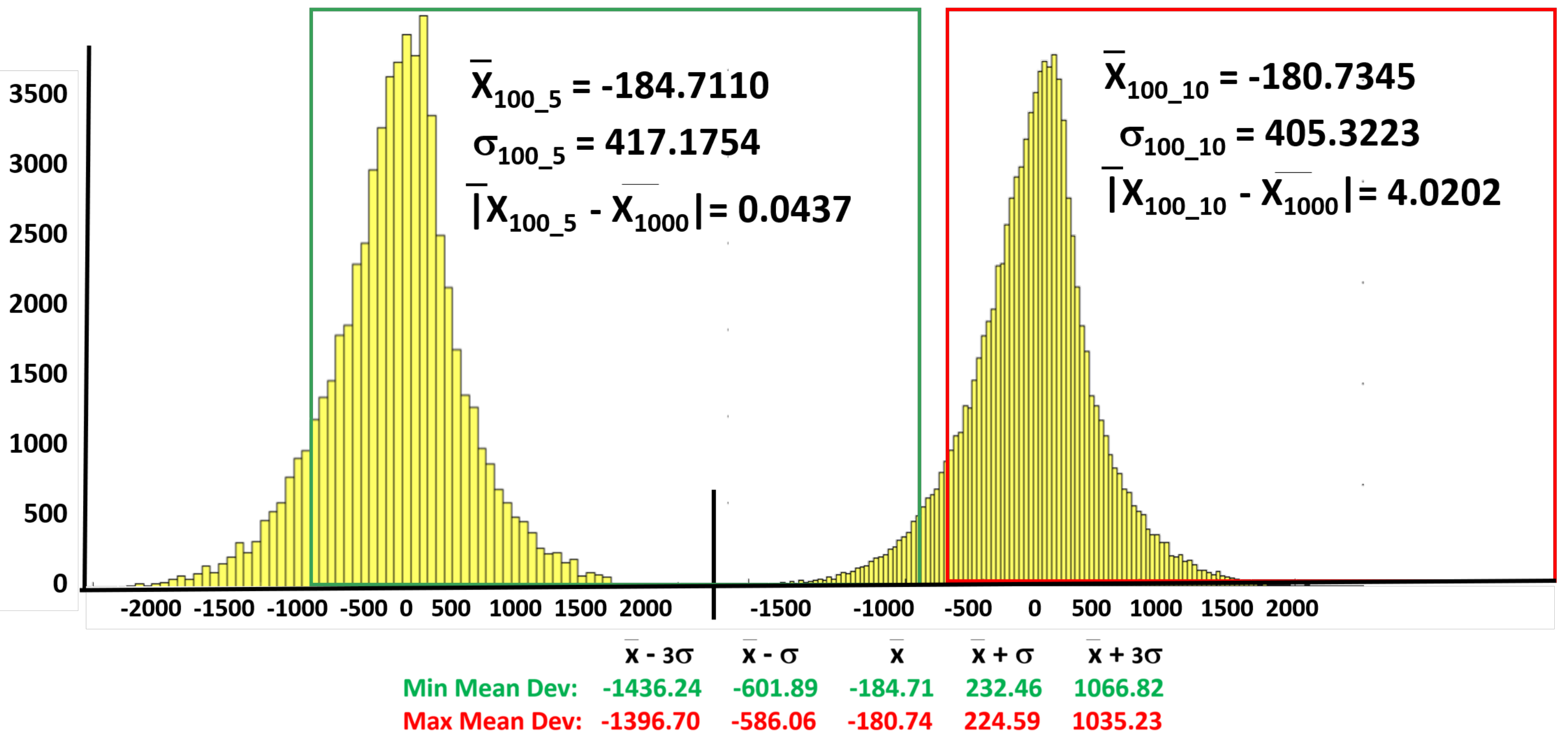}  
      \caption{Gaussian distribution, mean and standard deviation for 10x100 input vectors  showing plots with the maximum (in Red box) and minimum (in Green box) mean deviation of $3\sigma$ values of $conv1$ output for the 10x100 MNIST images compared with baseline (the maximum $3\sigma$ deviation is within $2.50\%$ of the baseline values). The full figure showing the distribution and statistical properties of all the 10 buckets each of 100 images is represented in Appendix\ref{appendix}.}
      \label{input_vectors1_d}
    \end{subfigure}
    \caption{Comparison of mean, standard deviation, Gaussian distribution and $3\sigma$ values of different sizes of datasets input images for LeNet CNN.}
    \label{input_vectors1}
\end{figure}

\begin{figure}[!t]
    \begin{subfigure}{\linewidth}
      \centering
      \includegraphics[width=0.5\linewidth]{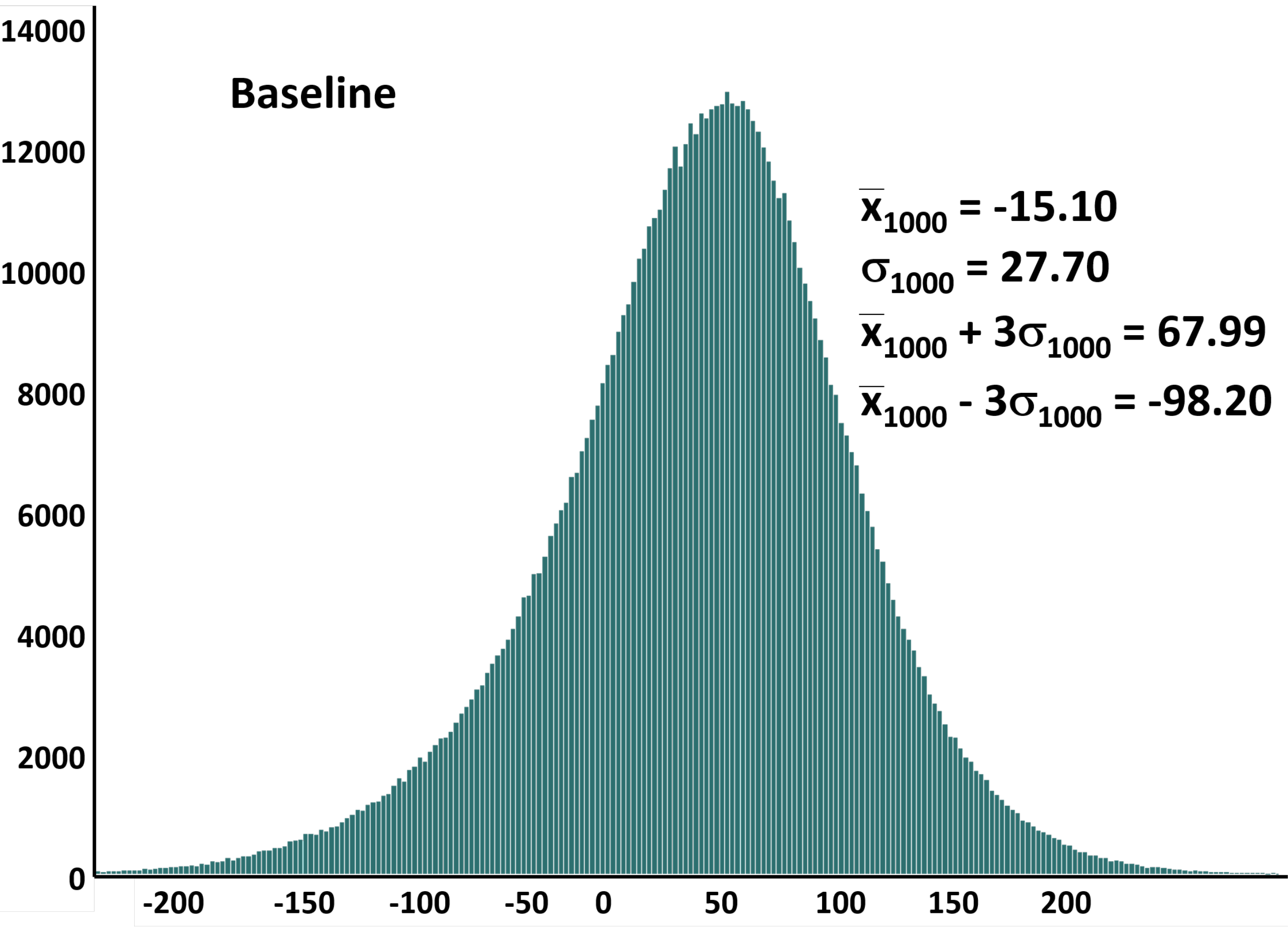}  
      \caption{Gaussian distribution, mean, standard deviation and $3\sigma$ values of $conv1$ output in response to 1000 input Cifar10 images (Baseline)}
      \label{input_vectors2_a}
    \end{subfigure}
    \begin{subfigure}{\linewidth}
      \centering
      \includegraphics[width=1\linewidth]{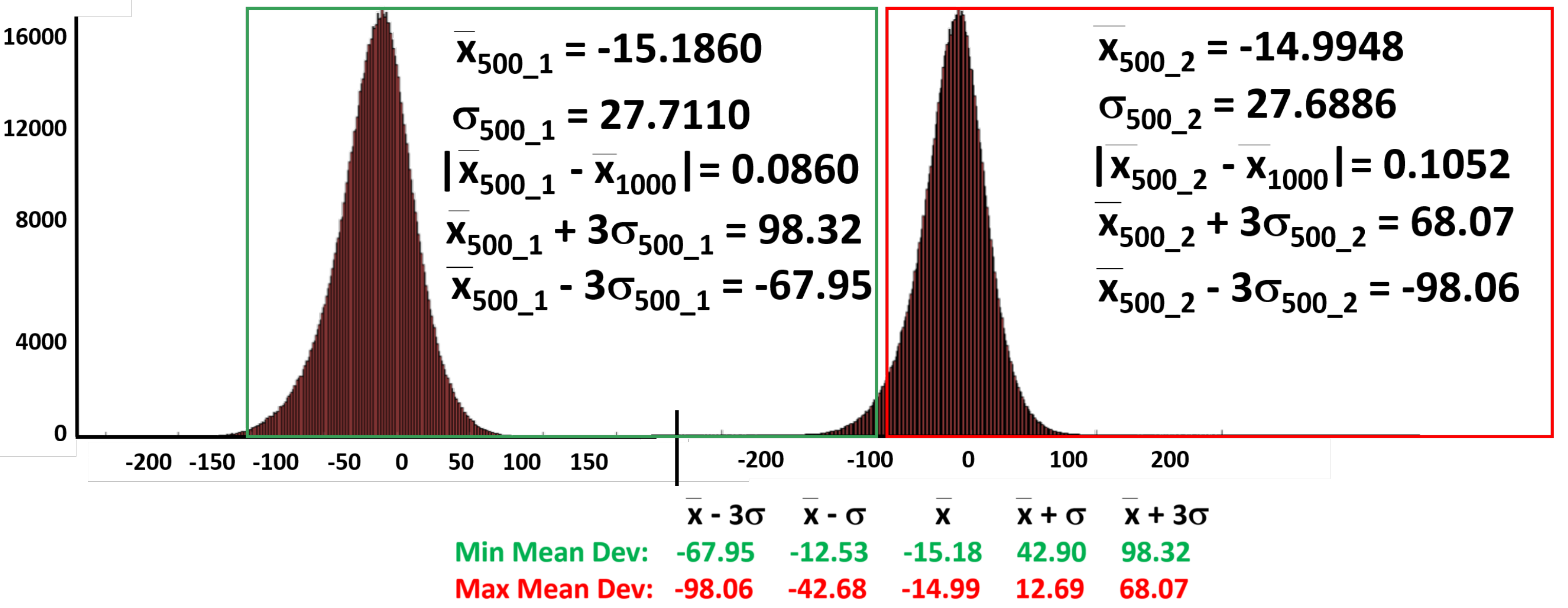}  
      \caption{Gaussian distribution, mean and standard deviation for 2x500 input vectors and also showing the maximum (in Red box) and minimum (in Green box) mean deviation of $3\sigma$ values of $conv1$ output for 2x500 Cifar10 images when compared with baseline (the maximum $3\sigma$ deviation is within $0.14\%$ of the baseline values). }
      \label{input_vectors2_b}
    \end{subfigure}
    \begin{subfigure}{\linewidth}
      \centering
      \includegraphics[width=1\linewidth]{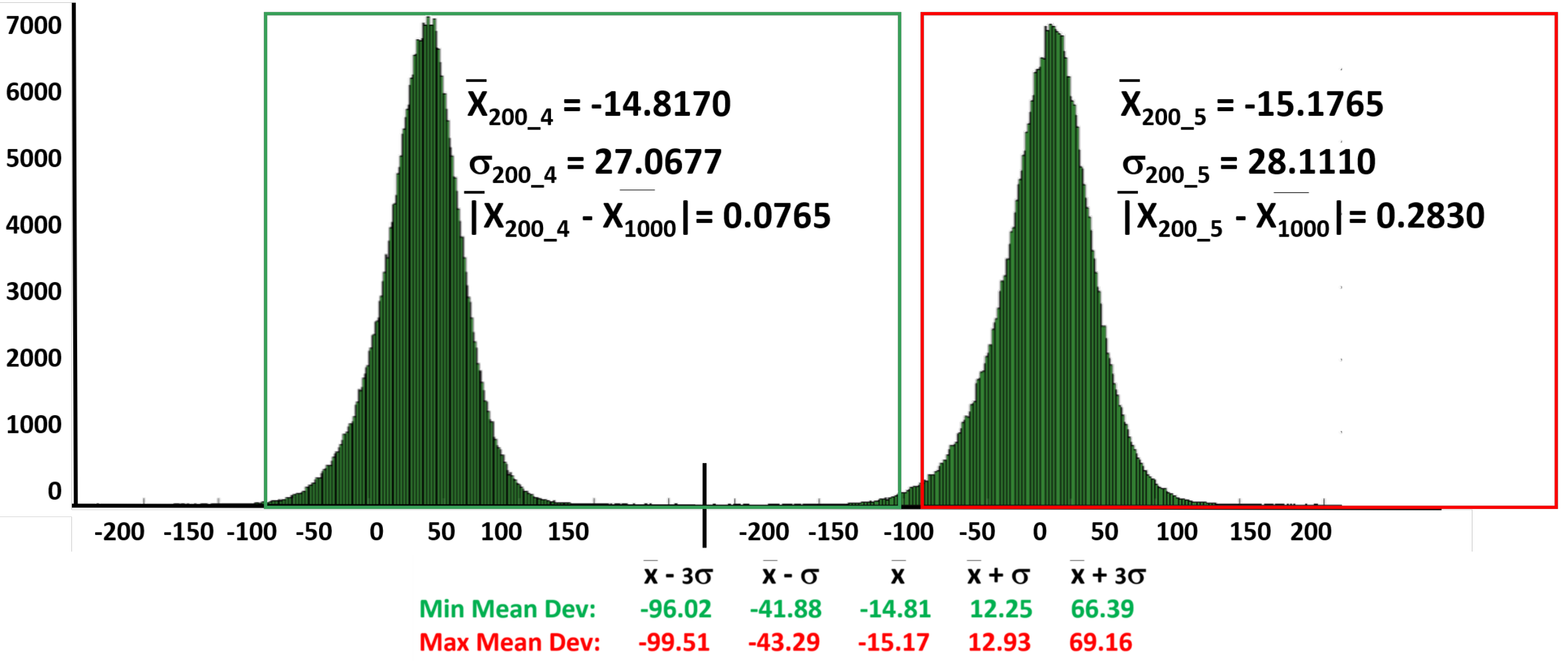}  
      \caption{Gaussian distribution, mean and standard deviation for 5x200 input vectors and also showing the maximum (in Red box) and minimum (in Green box) Mean deviation of $3\sigma$ values of $conv1$ output for 5x200 Cifar10 images compared with baseline. (the maximum $3\sigma$ deviation is within $2.36\%$ of the baseline values). The full figure showing the distribution and statistical properties of all the 5 buckets each of 200 images is represented in Appendix\ref{appendix}.}
      \label{input_vectors2_c}
    \end{subfigure}
    \begin{subfigure}{\linewidth}
      \centering
      \includegraphics[width=1\linewidth]{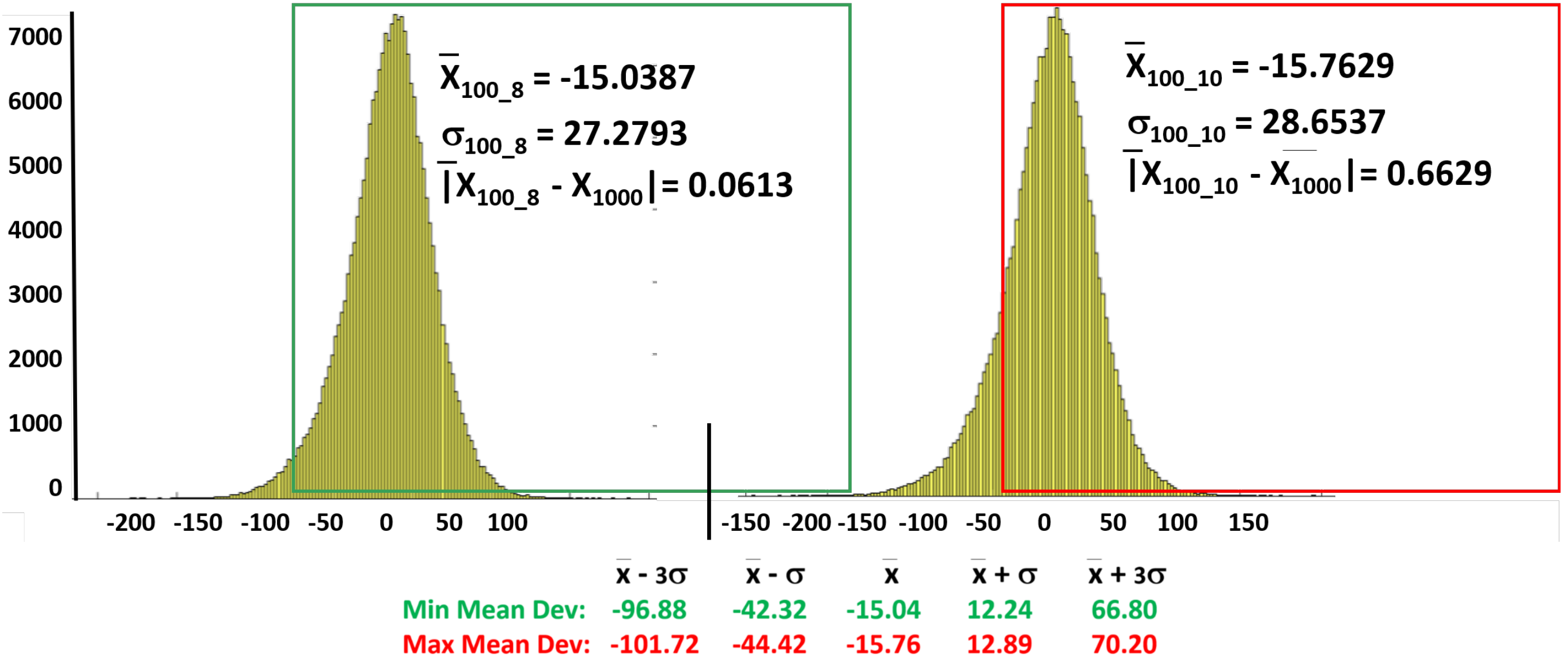}  
      \caption{Gaussian distribution, mean and standard deviation for 10x100 input vectors  showing plots with the maximum (in Red box) and minimum (in Green box) mean deviation of $3\sigma$ values of $conv1$ output for the 10x100 MNIST images compared with baseline (the maximum $3\sigma$ deviation is within $2.50\%$ of the baseline values). The full figure showing the distribution and statistical properties of all the 10 buckets each of 100 images is represented in Appendix\ref{appendix}.}
      \label{input_vectors2_d}
    \end{subfigure}
    \caption{Comparison of mean, standard deviation, Gaussian distribution and $3\sigma$ values of different sizes of datasets input images for LeNet CNN.}
    \label{input_vectors2}
\end{figure}

\begin{figure*}[t]%
    \centering
    \subfloat[Feature maps of $conv1$ layer is following Gaussian Distribution - Red Arrows indicate $3\sigma$ boundaries on either ends of the mean ]{{\includegraphics[width=5cm]{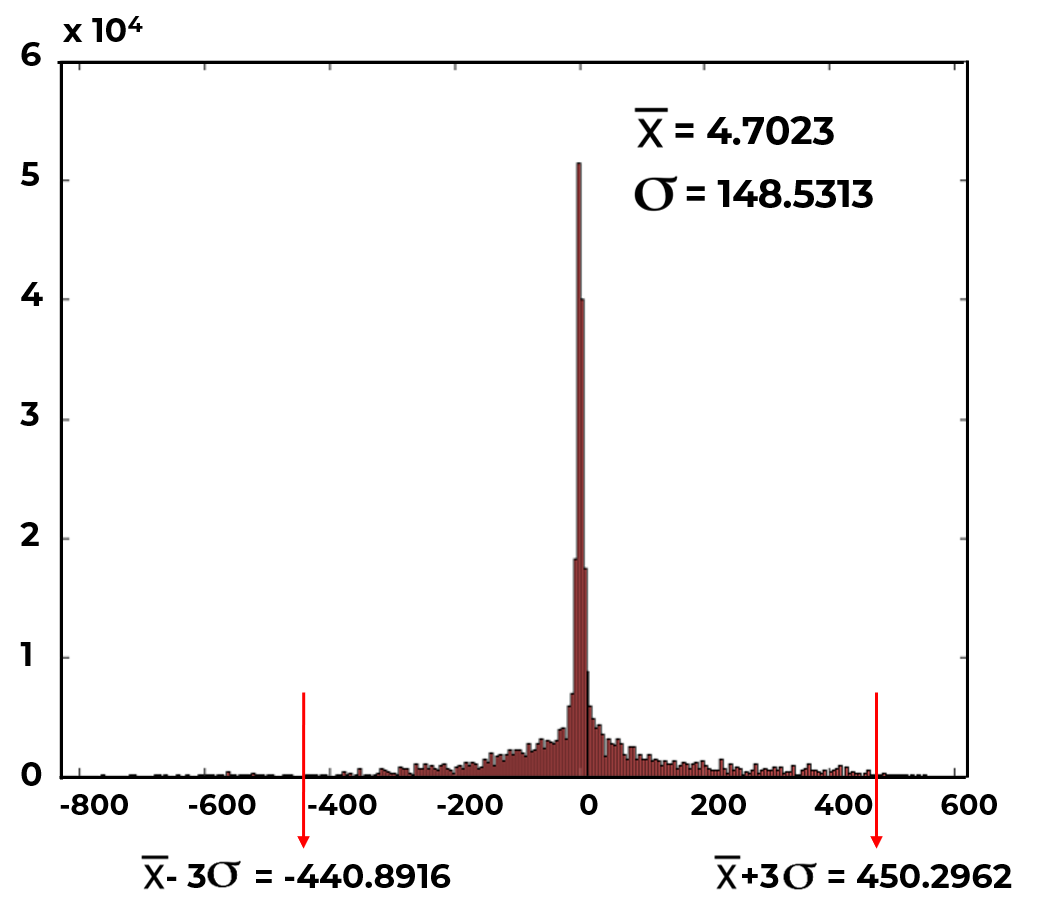} }}%
     \qquad
    \subfloat[Feature maps of $pool1$ layer is following Gaussian Distribution - Red Arrows indicate $3\sigma$ boundaries on either ends of the mean]{{\includegraphics[width=5cm]{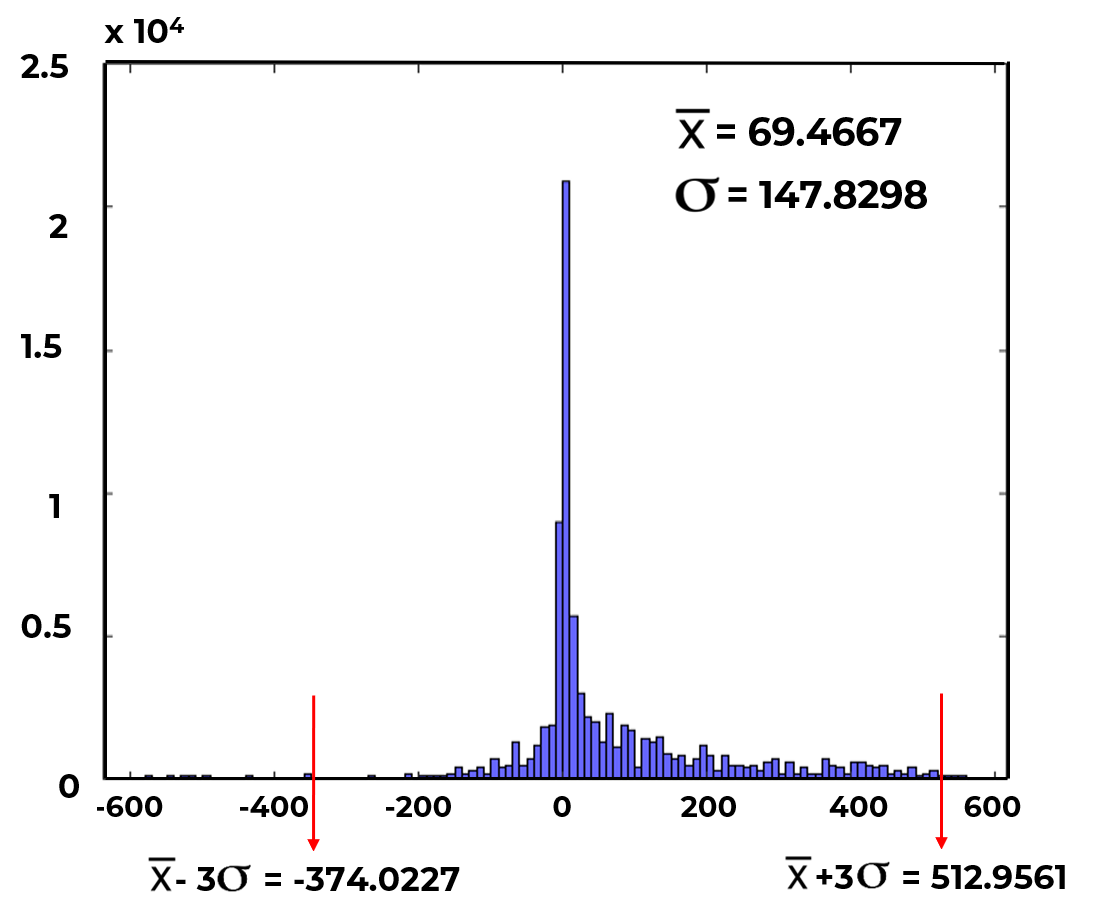} }}%
    \qquad
    \subfloat[Feature maps of $conv2$ layer is following Gaussian Distribution - Red Arrows indicate $3\sigma$ boundaries on either ends of the mean]{{\includegraphics[width=5cm]{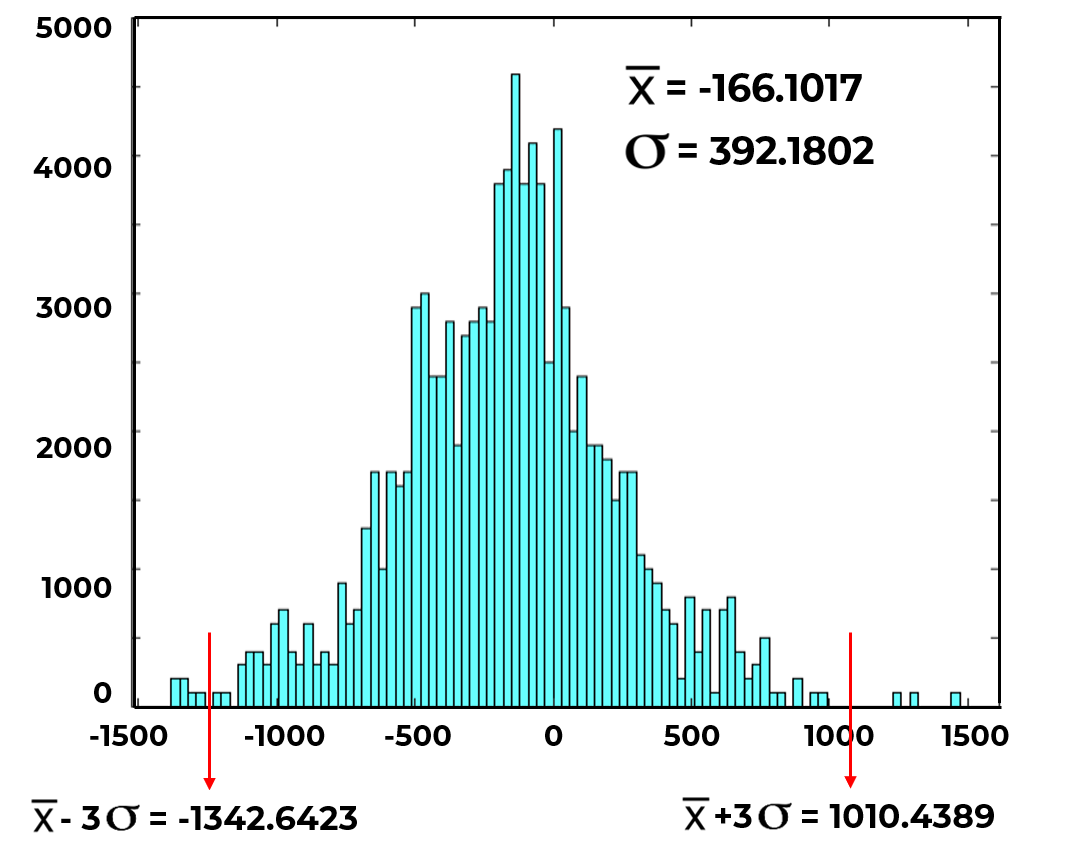} }}%
    \qquad
    \subfloat[Feature maps of $pool2$ layer is following Gaussian Distribution - Red Arrows indicate $3\sigma$ boundaries on either ends of the mean]{{\includegraphics[width=5cm]{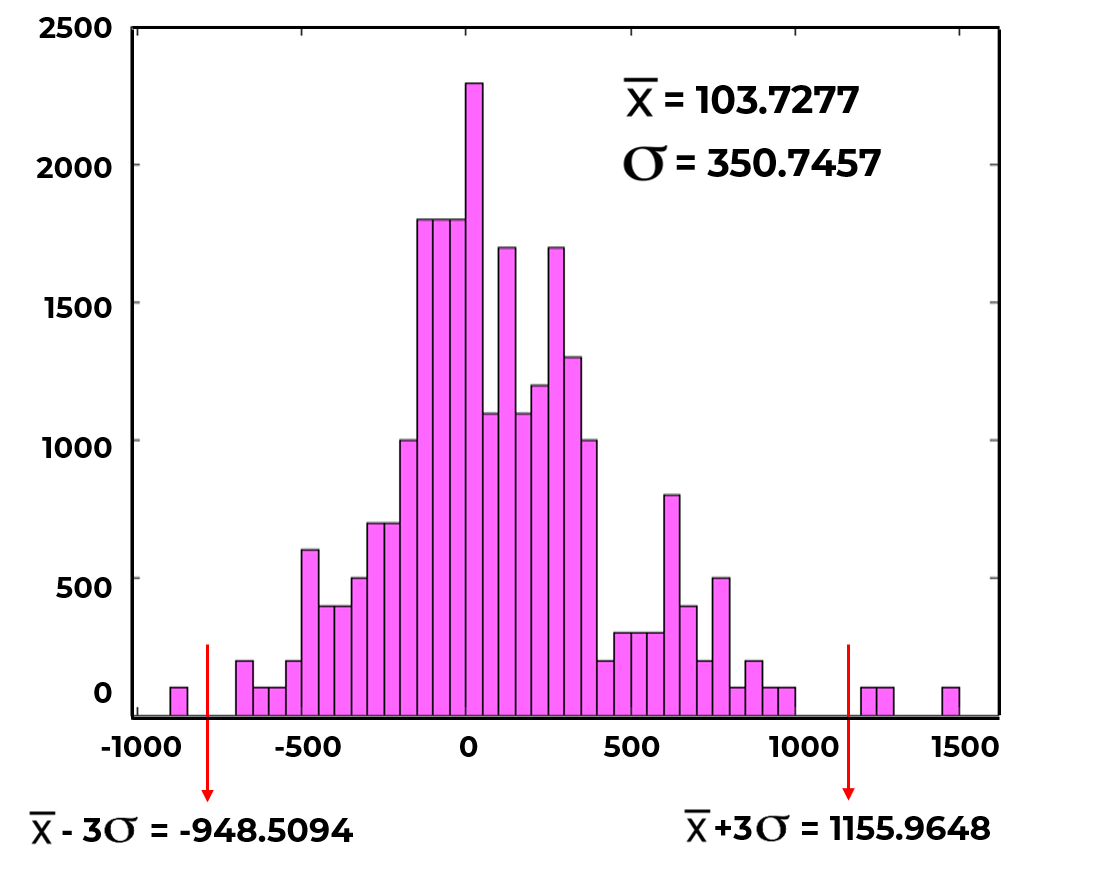} }}%
    \qquad
    \subfloat[Feature maps of $conv3$ layer is following Gaussian Distribution - Red Arrows indicate $3\sigma$ boundaries on either ends of the mean]{{\includegraphics[width=5cm]{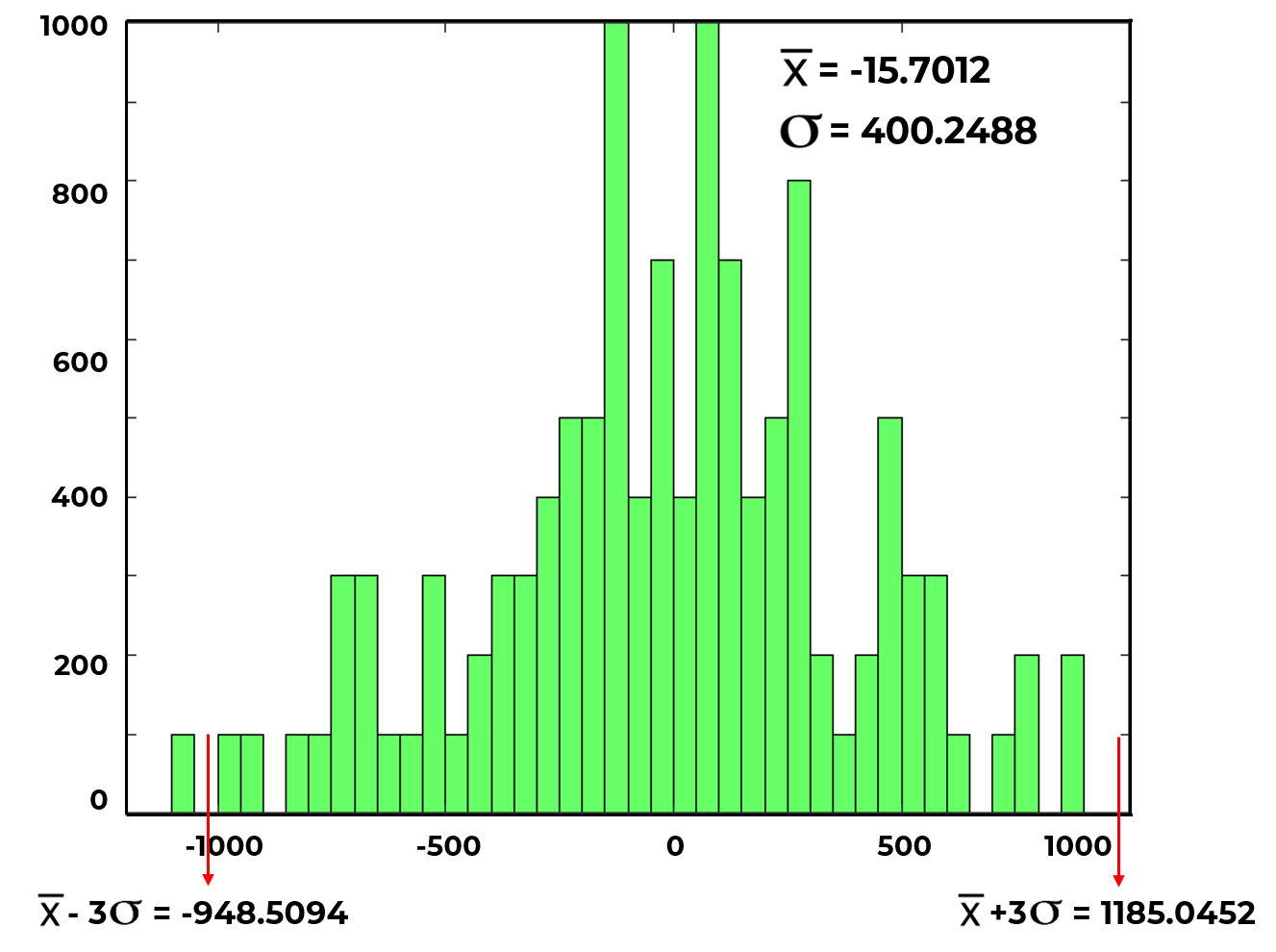} }}%
    \qquad
    \subfloat[Feature maps of $ip1$ layer is following Gaussian Distribution - Red Arrows indicate $3\sigma$ boundaries on either ends of the mean]{{\includegraphics[width=5cm]{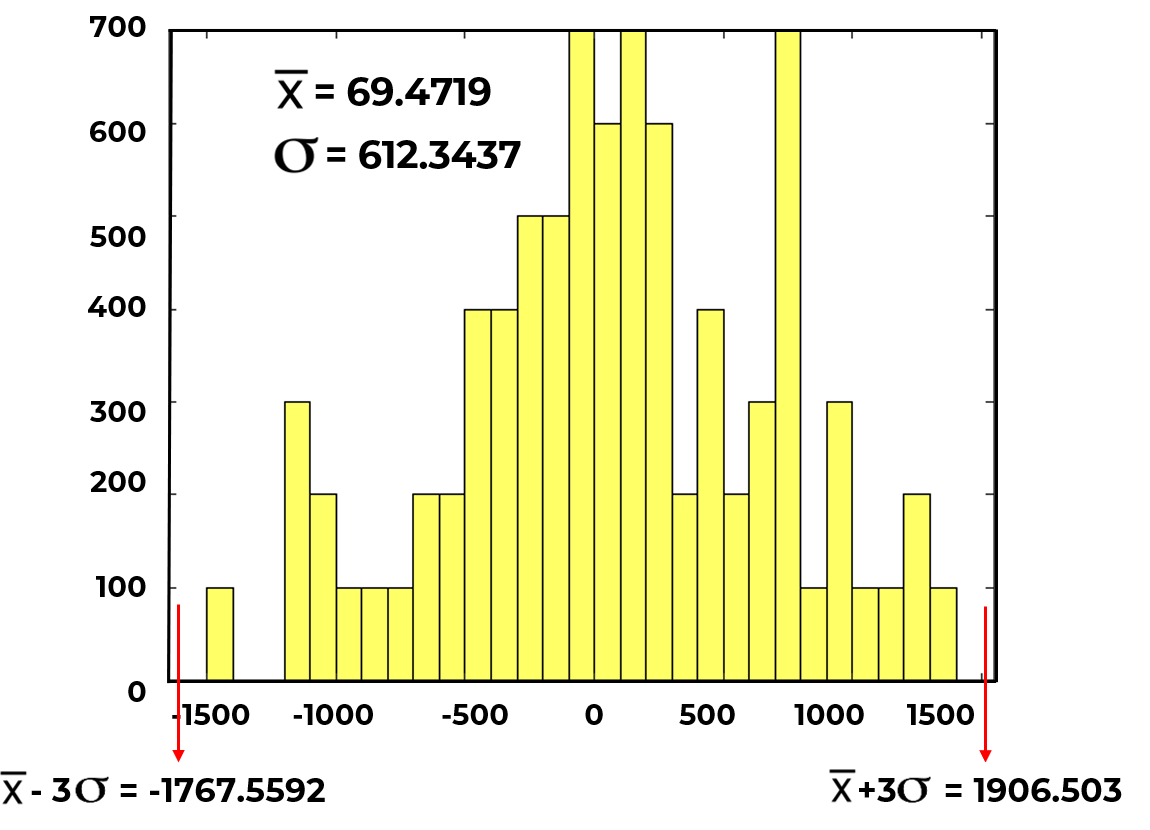} }}%
    \caption{This set of simulation indicates that output feature maps of each CNN layer, in response to minimal input data (100 images), shows close conformity to Gaussian Distribution for LeNet CNN}%
    \label{lenet_resulta}%
\end{figure*}

\subsection{Methodology Overview}
This section presents the step-by-step procedure of the proposed feSHI methodology to design the static and dynamic hardware attacks for multi-FPGA-based CNN inference.
\subsubsection*{Validation Dataset Sample Space}
\label{size of dataset}
For the offline pre-processing phase, to obtain stealthy TBCs, we require FMV-Dataset. Figs~\ref{input_vectors1} and~\ref{input_vectors2} show a systematic approach employed in obtaining statistical attributes from a limited FMV-Dataset made available to the 3P designer. This systematic approach involves the analysis of different dataset sizes in terms of the relationship between their feature maps and their statistical distribution. From the analysis, we obtain stealthy TBCs inserted in the hardware design to serve as the trigger. Gaussian distribution plots of the aggregated output feature maps of the CNN layers can be exploited for the FeSHI hardware attack. Figs.~\ref{input_vectors1} and~\ref{input_vectors2} show output feature maps of the $conv1$ layer of LeNet and LeNet-3D CNN models for the MNIST and Cifar10 datasets respectively (all the graphs of statistical properties of the 5$\times$200 datasets and 10$\times$100 dataset are shown in Appendix I). MNIST dataset consists of 60,000 training images and 10,000 testing images \cite{lecun1998gradient}. We consider 10\% of the testing images (1000 input vectors) as the baseline dataset. In order to find out minimal sample space to serve as the size of the FMV-Dataset, which can provide the attacker with a good estimate of the statistical properties, the baseline input vectors are divided into different subsets. The respective statistical properties (mean and standard deviation) of the different subsets (the 1000 input vectors are randomly divided into different buckets of $2\times500$, $5\times200$, $10\times100$) of the 1000 input vectors  (FMV-Dataset) and compared with the properties of the baseline input vectors. 

The approach shown in Figs.~\ref{input_vectors_appendix1} and~\ref{input_vectors_appendix2} in Appendix 1 (Section~\ref{appendix}) provides an avenue of finding appropriate RoVs from seen data (FMV-Dataset) that can be utilized as TBC during the inference phase applied for real-time classification on unseen data. The Gaussian distribution, $3\sigma$ values of the subset buckets of the dataset showing the minimum and maximum mean deviations from the baseline of all the reduced datasets are compared with the baseline dataset and shown in Figs. ~\ref{input_vectors1} and ~\ref{input_vectors2}. The RoVs are chosen from values that lie beyond the $-3\sigma$ values on either side of the mean of the Gaussian distribution. From the Gaussian distribution plots in Figs.~\ref{input_vectors1}, the $3\sigma$ regions of the plots with minimum mean deviation (across the $2\times500$, $5\times200$ and $10\times100$ test vectors for LeNet) show highest percentage differences of $4.50\%$, $1.40\%$, and $2.50\%$ respectively for the $3\sigma$ values in comparison with the baseline plot. From Fig.~\ref{input_vectors2}, the plots with maximum mean deviation (across the $2\times500$, $5\times200$ and $10\times100$ test vectors for LeNet-3D) show highest percentage differences of $0.14\%$, $2.36\%$, and $3.59\%$ respectively for the $3\sigma$ values in comparison with the baseline plots. Our results stated above validates our Hypothesis 1, which stipulates that we can estimate the statistical properties of a baseline dataset from a limited FMV-Dataset. This is used to successfully obtain stealthy TBCs and design dynamic triggers discussed in Section~\ref{dynamic_attack1}.
\begin{figure*}[t]
    \centering
    \includegraphics[width=1\linewidth]{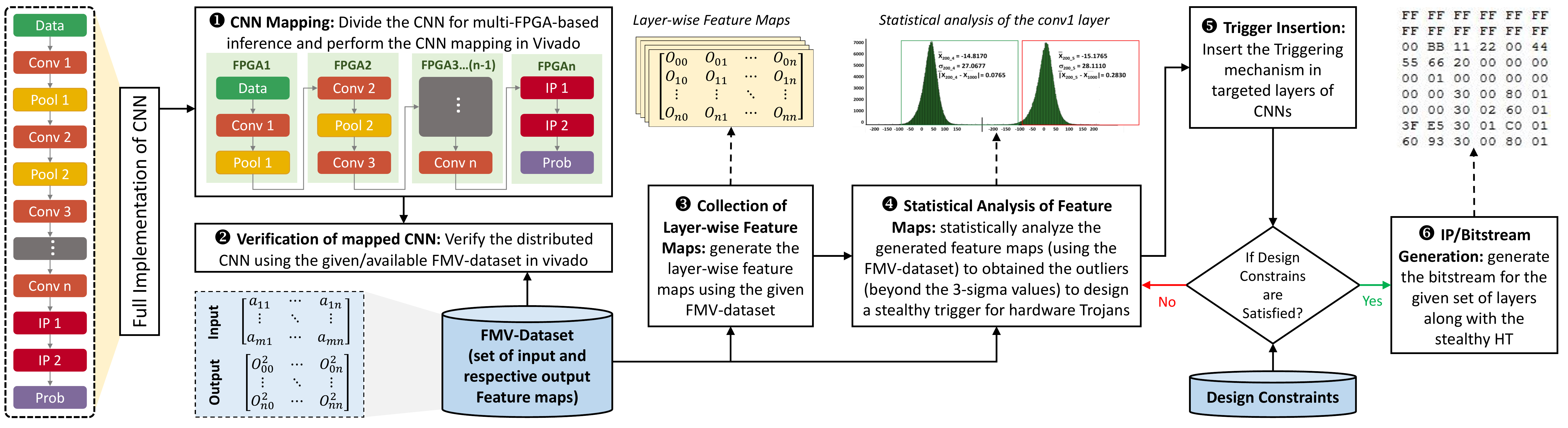}
    \caption{Proposed FeSHI methodology offline overview: The offline processing of the hardware Trojan showing the $3\sigma$ analysis, extraction of the triggering boundary conditions (TBC) and insertion of the hardware Trojan}
    \label{overview_off}
\end{figure*}

The FeSHI methodology is divided into 2 stages, namely: the offline pre-processing stage and the run-time stage. An overview of the proposed Offline pre-processing stage is shown in Fig.~\ref{overview_off}. This stage provides a means of selecting one or more stealthy TBCs to serve as the trigger(s) for the attack. These trigger(s) can be targeted at one or more layers depending on the choice of the adversary. This is further elaborated on in Section~\ref{trigger}. The run-time operation discusses the resulting effect once the hardware Trojan is triggered, which is elaborated in Section~\ref{payload}.

\section{Offline pre-processing: TBC to Design Triggers}
\label{trigger}
During the verification stage using the FMV-Dataset, the adversary keeps track of the output feature maps of the CNN layers available, as shown in Fig.~\ref{overview_off}. As discussed in Section~\ref{attack_meth}, due to the standard weight initialization approaches and innate random nature of incoming features, the computation results in an output feature map that follows Gaussian distribution. With this understanding, the attacker can obtain a particular RoVs (subsequently called Triggering Boundary Condition (TBC)) that lies beyond the $\pm3\sigma$ values of the Gaussian distribution (as shown in Line $7$ of Algorithm~\ref{alg:MYALG_static} and explained in further detail in Sections~\ref{static_attack1} and~\ref{dynamic_attack1}). The TBCs infrequently occur, as seen in Fig.~\ref{overview_off} and represent less than $1\%$ of the total output feature map values across the FMV-Dataset.

To validate that the output feature maps of CNN layers follow a Gaussian distribution, we analyzed them by plotting a histogram of their aggregated numerical values, as illustrated in Fig.~\ref{lenet_resulta}. The histogram plots of the aggregated feature maps of each individual layer (from $conv1$ to $ip1$ layers) of LeNet CNN in response to $100$ uniformly distributed images (serving as the FMV-Dataset) are obtained and plotted in Fig.~\ref{lenet_resulta}. The output feature maps, as expected, is having close conformity to  Gaussian distribution from where an attacker can obtain the rare TBCs (e.g., for the case of $conv1$, the value can be greater than $450.2962$ or less than $-440.8916$, as indicated by the red arrows). These boundary conditions are obtained and used to design two types of attacks, namely: Static attacks and dynamic attacks. These are further explained next. 
\subsection{Static Attack}
\label{static_attack1}
\begin{figure}[h]
    \centering
    \includegraphics[width=1\linewidth]{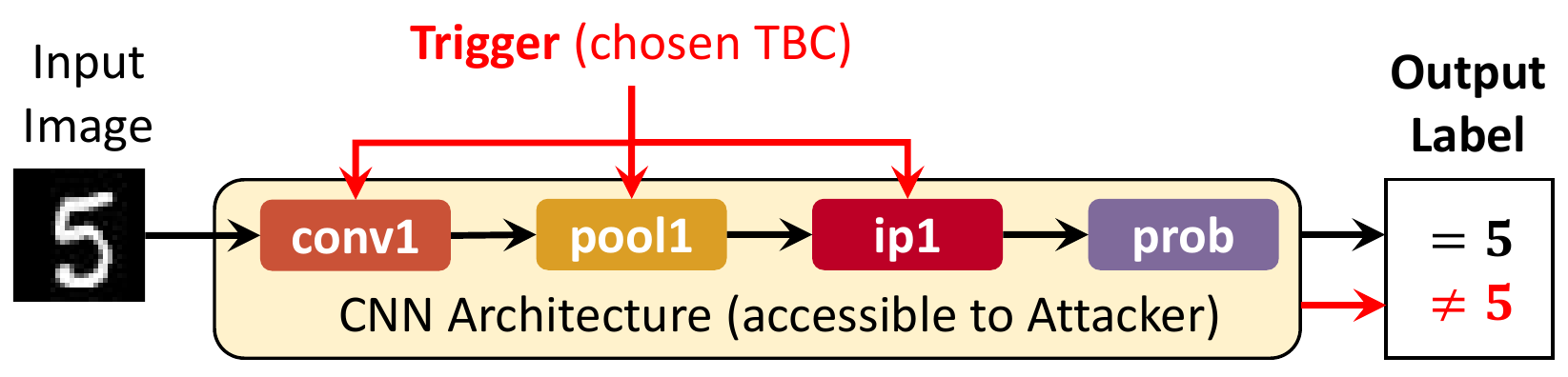}
    \caption{Static Attack design showing the insertion of a TBC trigger to the layer of the CNN available to the attacker during the runtime}
    \label{static}
\end{figure}

In this attack, the attacker aggregates the feature maps generated in response to the FMV-Dataset (as depicted in Line $4$ of Algorithm~\ref{alg:MYALG_static}). The Gaussian distribution of the aggregated feature map is plotted to observe the $3\sigma$ values. One of the RoVs (as shown in Line $6-9$ of Algorithm~\ref{alg:MYALG_static}) that lie beyond the $3\sigma$ values is selected and inserted into the hardware design to serve as the TBC for the attack to the targeted CNN layer available to the attacker as shown in Fig.~\ref{static}. For the proposed design, \textbf{the attack is evaluated based on the stealthiness (additional hardware overhead  (BRAM,  DSP,  flip-flops (FFs),  look-up tables (LUTs), Latency) and effectiveness (the rate of triggering) of the attack}. Lines $9-18$ of Algorithm~\ref{alg:MYALG_static} evaluates the selected TBC in terms of the degree of occurrence of the trigger (effectiveness) by testing it on the feature map of the targeted CNN layer using the FMV-Dataset (seen data). If the amount of triggering satisfies some preset number set as the threshold amount of trigger occurrence, as seen in Line $15$ of Algorithm~\ref{alg:MYALG_static}, then this RoV is set as the TBC. In this work, to achieve a maximum of 10\% triggering probability, the threshold is set to 25 (number of occurrences). If the threshold condition is not satisfied, then another RoV that lies beyond the $3\sigma$ values will be chosen, and testing the threshold condition will be repeated. The threshold can be changed based on design choice in design time.

To achieve a resource-friendly design trade-off between resource usage and the number of possible triggers (that will lead to reduced accuracy), we provided three different variants of this attack which are discussed in Section~\ref{attack variation} (Lines 20 - 26 in Algorithm~\ref{alg:MYALG_static}).

\subsection{Dynamic Attack}
\label{dynamic_attack1}
In a situation where the defender tries to come up with test vectors to detect hardware attacks due to such rarely occurring events, we propose a dynamic version of the FeSHI attack that changes its RoVs to boost stealthiness during runtime.

\begin{figure}[h]
    \centering
    \includegraphics[width=1\linewidth]{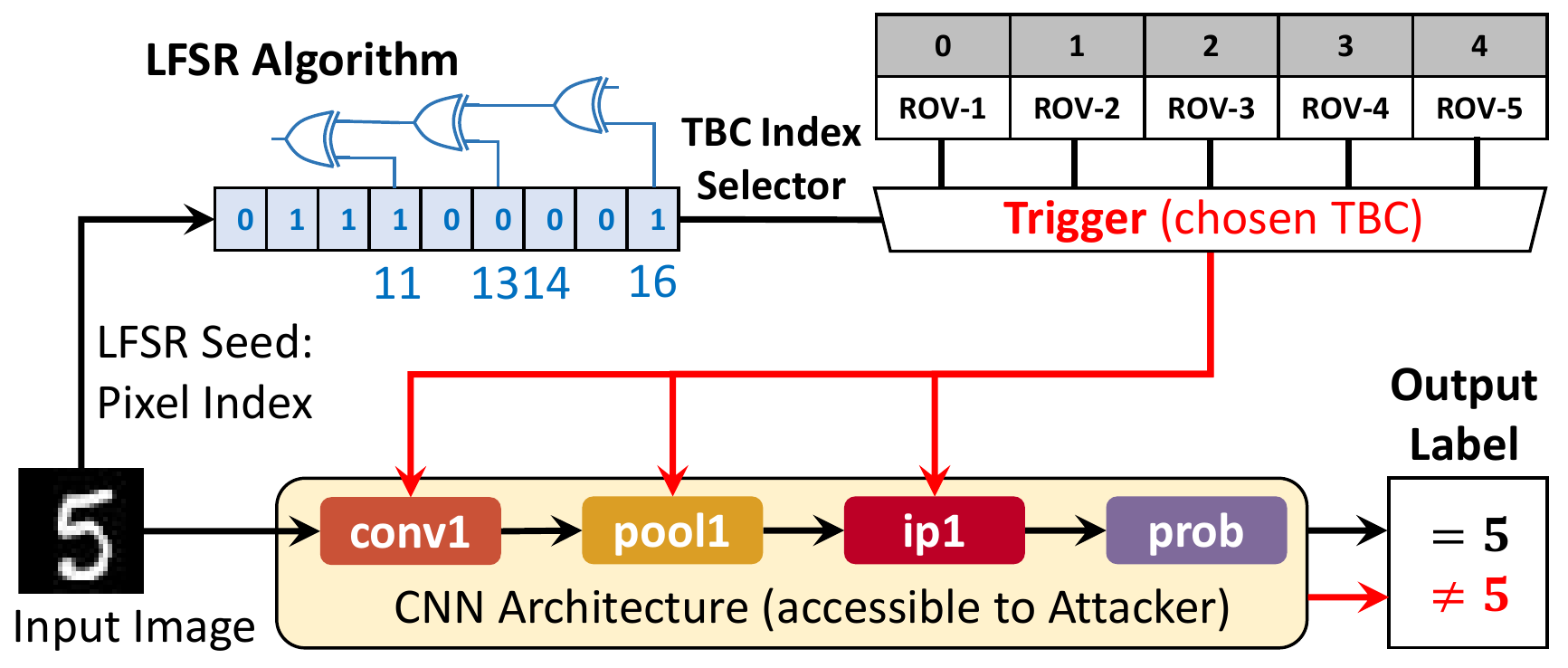}
    \caption{Conceptual depiction of FeSHI dynamic attack design showing the use of LFSR algorithm to implement different TBC trigger for different image instances- This may randomly choose one of the five RoV as a TBC, hence it creates another layer of illusion for such attacks to be detected using conventional test vector techniques.}
    \label{dynamic}
\end{figure}

In this attack, after collection and aggregation of the feature maps, multiple sets of RoVs that lie beyond the $3\sigma$ values of the Gaussian distribution of the aggregated feature maps are inserted as TBCs (depicted in lines $3 - 9$ of Algorithm~\ref{alg:MYALG1_dynamic}) in the hardware design in the form of an array (TBC array) as depicted in Fig.~\ref{dynamic}. In this work, an image cycle refers to every time an input image or feature map is passed as input for classification during runtime. From Fig.~\ref{dynamic} during runtime, in an image cycle, the value present in a pre-selected index of an input image is normalized and serves as a seed to the LFSR (Linear Feedback Shift Registers) algorithm. LFSR algorithms are used for generating non-sequential sets of numbers in a time-efficient way by making use of right-shift, and XOR operations \cite{Maxim2010}. 

The LFSR used in this work is a 6th-degree polynomial ($q^6 + q + 1$). In this work, the output of the LFSR algorithm is random integers between $0 - 4$ in order to satisfy the selection of one of the $5$ RoVs selected and inserted in the hardware design. The LFSR algorithm generates a TBC index selector (designated as $selector$ in Algorithm~\ref{alg:MYALG1_dynamic}) that lies within the length of the TBC array, which serves as the select bit to the TBC array (array of multiple RoVs) multiplexor  (as shown in Line $12$ of Algorithm~\ref{alg:MYALG1_dynamic}). Based on the TBC index selector, one of the RoVs in the TBC array is selected and inserted into the CNN layer to be attacked. This is depicted in Lines 11 - 16 of Algorithm~\ref{alg:MYALG1_dynamic}. Here, $5$ sets of RoVs are selected for the TBC array in order to increase randomness, improve stealthiness, and also incur minimum additional resources. The LFSR algorithm is used to randomly generate values from $0$ to $4$, serving as the TBC array index selector. The RoVs in  TBC array corresponding to the index selected serve as the TBC trigger in the current image cycle. Lines $11-19$ of Algorithm~\ref{alg:MYALG1_dynamic} evaluates the stealthiness of the  LFSR algorithm and the selected RoVs serving as the TBC array. If the effectiveness satisfies the triggering threshold, the RoVs are selected. Lines $22-28$ of Algorithm~\ref{alg:MYALG1_dynamic} shows the selection of appropriate attack variation approach in terms of resource utilization and attack variation in terms of layer-wise, channel-wise, or index-wise approach. This is discussed in the next subsection.

\begin{table*}[]
    \centering
    \caption{Chart comparing all of FeSHI attack variations in terms of Resources, Latency and Triggering Probability}
    \label{chart_attack_vary}
    \resizebox{1\linewidth}{!}{
    \begin{tabular}{|c|c|c|c|}
    \hline
    \begin{tabular}[c]{@{}c@{}}\textbf{Attack}\\ \textbf{Variations}\end{tabular} & \textbf{Number of Resources} & \textbf{Latency (Clock-Cycles)} & \textbf{Triggering Probabilities} \\ \hline
    \textbf{Layer-wise} & \begin{tabular}[c]{@{}c@{}}Highest (Requires more resources to \\ compare \textit{all elements of FM}   with TBC)\end{tabular} & \begin{tabular}[c]{@{}c@{}}Highest (Requires more time to compares \\ \textit{all elements of the FM} with TBC)\end{tabular} & {\begin{tabular}[c]{@{}c@{}}Highest (This compares \textit{all} \\ \textit{elements of the FM} with TBC)\end{tabular}} \\ \hline
    \textbf{Channel-wise} & {\begin{tabular}[c]{@{}c@{}}Medium (Requires fewer resources to \\ compare \textit{a channel of the FM} with TBC)\end{tabular}} & {\begin{tabular}[c]{@{}c@{}}Medium(Requires lower time to compare  \\ \textit{a channel of the FM} with TBC)\end{tabular}} & {\begin{tabular}[c]{@{}c@{}}Medium(Only compares \textit{a channel}\\  \textit{of the FM} with TBC)\end{tabular}} \\ \hline
    \textbf{Index-wise} & {\begin{tabular}[c]{@{}c@{}}Lowest (Requires less resources to \\ compare \textit{one value of the FM} with TBC)\end{tabular}} & {\begin{tabular}[c]{@{}c@{}}Lowest (Requires least amount of time to \\ compare \textit{one value of the FM} with TBC)\end{tabular}} & {\begin{tabular}[c]{@{}c@{}}Lowest(Only compares \textit{one value}\\ \textit{of the FM} with TBC)\end{tabular}} \\ \hline
    \end{tabular}}
\end{table*}

\begin{figure}[]
    \centering
    \includegraphics[width=1\linewidth]{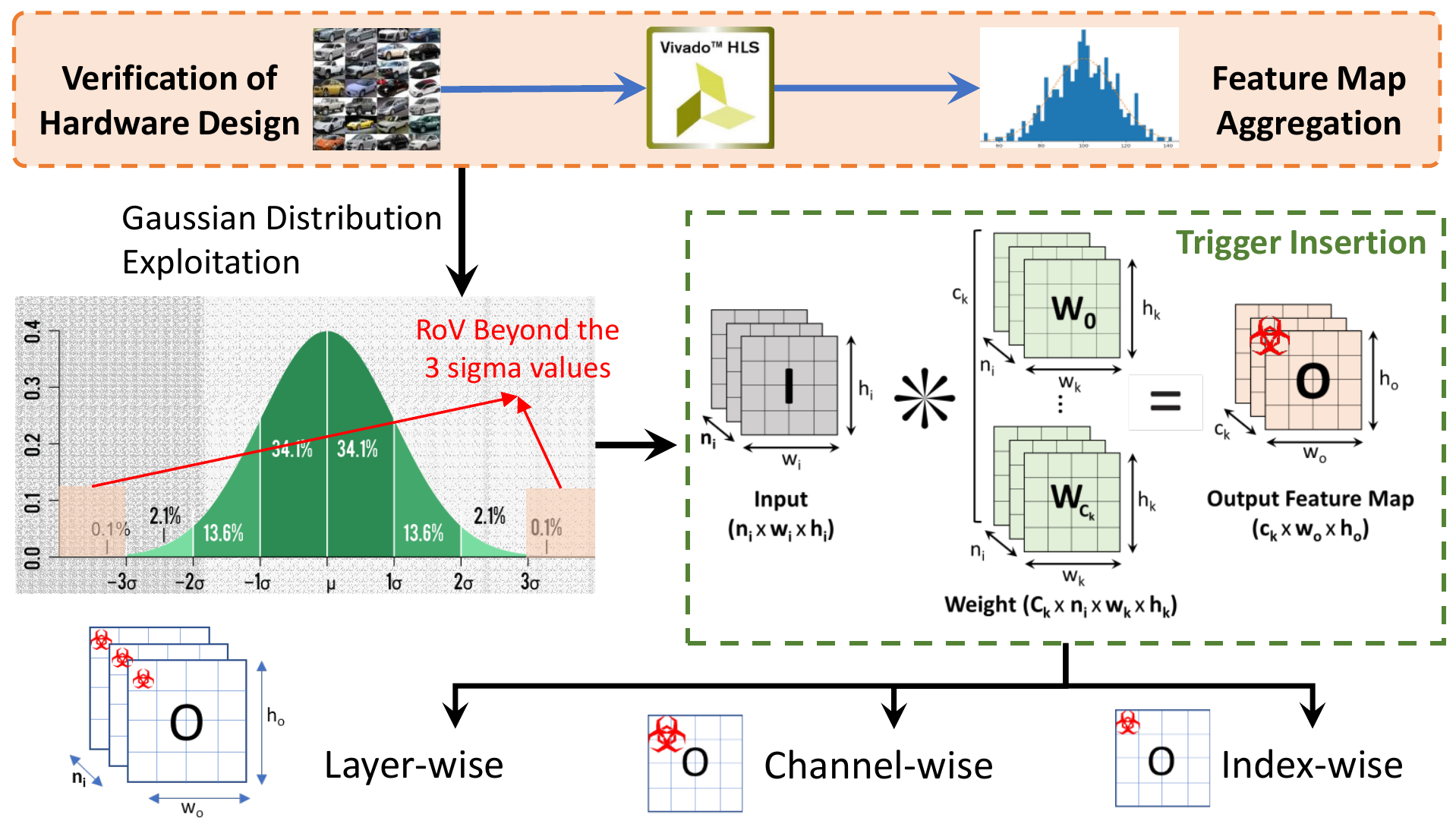}
    \caption{FeSHI offline trigger variation showing the insertion of $3\sigma$ values in either the layer (all of the channels of the feature map are compared to the selected TBC(s)) or channel (any one channel ($x$) of the feature map is compared to the selected TBC(s)) or a particular index (just one value of the feature map are compared to the selected TBC(s)) of the feature maps}
    \label{offline11}
\end{figure}

\subsection{Attack Trigger Variations}
\label{attack variation}
Both static and dynamic attacks are implemented in three different variations of trigger design. These different variations of triggers show different design trade-offs between stealthiness, resources, and latency. Algorithms~\ref{alg:MYALG_static} and~\ref{alg:MYALG1_dynamic} consider these trade-offs and can provide a more resource-friendly solution for a given context (line 20 in Algorithm~\ref{alg:MYALG_static}, and line 22 in Algorithm~\ref{alg:MYALG1_dynamic}).

\subsubsection{Layer-wise Trigger Design: Low Stealth/High Resource}
In this trigger design, all the elements across all channels of the targeted CNN layer output feature map of interest are compared against the embedded TBC as shown in Fig.~\ref{offline11}. If one or more of the values of the feature maps fall within the TBC, the payload gets activated. The probability of trigger getting activated is highest at the expense of high resources, see Table~\ref{chart_attack_vary} for comparison. 

\subsubsection{Channel-wise Trigger Design: Medium Stealth/Medium Resource}
In this case, all the elements of a channel of the targeted CNN layer output feature maps are compared against the predefined TBC as depicted in Fig.~\ref{offline11}. If one or more elements in the channel satisfy the TBC, the payload gets activated. As shown in Table ~\ref{chart_attack_vary}, it uses lesser resources than the layer-wise trigger, but consequently, the probability of getting triggered is also lesser.

\subsubsection{Index-wise Trigger Design: High Stealth/Low Resource}
In this case, only an index in the targeted CNN layer output feature map is monitored and compared with the predefined TBC as shown in Fig.~\ref{offline11}. If the value of the chosen index satisfies the TBC, the payload gets activated. This requires the least resources and consumes the least latency, but also triggering probability is also the least among all the variants of this attack as stated in Table ~\ref{chart_attack_vary}.

From Fig.~\ref{offline11}, we consider a scenario where a convolution layer with an output feature map of shape $6 \times 24 \times 24$ is exposed to the layer-wise trigger, all the elements ($3456$ elements) in the output feature map are tested against the triggering condition. Hence $3456$ elements have the likelihood of triggering the attack. If the aforementioned convolution layer is exposed to the channel-wise attack, all the elements in just one channel (of shape $1 \times 24 \times 24$) of the layer are tested against the selected triggering conditions. Hence only $576$ elements have the likelihood of triggering the attack. In index-wise trigger design, only one element of the output feature map is tested against the selected triggering condition. Hence in an index-wise attack, just one element has the likelihood of triggering the attack. Hence layer-wise attack requires some sort of computation done to all the elements ($3456$ elements) in the output feature map compared to $576$ elements in the channel-wise attack and $1$ element in the index-wise attack. Hence, the possibility of triggering simply increases due to more number of computations done in a Layer-wise attack. 

\begin{algorithm}[]
\footnotesize
\captionsetup{font={sf,footnotesize},labelfont={bf,color=accessblue}}
\caption{\footnotesize Offline pre-processing: Static Attack where: DC: Design Constraints, LW: Layer-wise, CW: Channel-wise}
    \begin{algorithmic} [1]
        \REQUIRE Mapping of the CNN in HLS (C++).
        \REQUIRE Verification of design
        \STATE Identify targeted layer $L$
        \STATE $trig = 0$ (number of triggers)
        \FOR {image ($X_i$) $\in$ FMV-Dataset}
        \STATE Collect $AF = \{O_1, O_2,..., O_{no}\}$\\
        Where: $O_i,...,O_{no} \in O$ (Values in the feature maps)\\
        $n_{o}$ is the maximum amount of channels in the feature maps\\
        $AF$ represent aggregated feature maps\\
        \ENDFOR
        \STATE Obtain values beyond $3\sigma$ values in $AF$.
        \STATE Collect low occurring RoV $[a, b]$ $\in$ $AF$\\
        where: $a$ represent the lower limit of the selected RoV\\
        $b$ represent the upper limit of the selected RoV\\
        \STATE Insert $[a, b]$ in Hardware design
        \STATE Test $[a, b]$
        \FOR {image ($X_i$) $\in$ FMV-Dataset}
        \IF {$(\{O_i,...,O_z\} \in [a, b])$}
        \STATE $trig += 1$
        \ENDIF
        \ENDFOR
        \IF {$trig <= threshold$}
        \STATE Insert $[a, b]$ as TBC
        \ELSE
        \STATE Repeat steps 3 to 16
        \ENDIF\\
        where: $threshold$ represent the desired maximum amount of times a number in the feature map of a layer falls in the seleted RoVs in response to the FMV-Dataset\\
        \IF {TBC $\in$ DC as LW trigger}
        \STATE Insert TBC as LW trigger
        \ELSIF {TBC $\in$ DC as CW trigger}
        \STATE Insert TBC as CW trigger
        \ELSE
        \STATE Insert as index-wise trigger
        \ENDIF\\
        \end{algorithmic}
    \label{alg:MYALG_static}
\end{algorithm}

\begin{algorithm}[]
\footnotesize
\captionsetup{font={sf,footnotesize},labelfont={bf,color=accessblue}}
\caption{Offline pre-processing: Dynamic Attack where:
DC: Design Constraints
LW: Layer-wise
CW: Channel-wise}
\begin{algorithmic} [1]
\REQUIRE Mapping of the CNN in HLS (C++).
\REQUIRE Verification of design
\STATE Identify targeted layer $L$
\STATE $trig = 0$
\FOR {image ($X_i$) $\in$ FMV-Dataset}
\STATE Collect $AF = \{O_1, O_2,..., O_{no}\}$
\ENDFOR
\STATE Obtain values beyond $3\sigma$ values in $AF$.
\STATE Collect $5$ low occurring RoVs $\{[A_0, B_0]$, $[A_1, B_1]$, $...$, $[A_z, B_z]\}$ $\in$ $AF$
Where: $A_0 - A_z$ represent the upper limits of the selected ROVs\\
$B_0 - B_z$ represent the lower limits of the selected ROVs\\
$z$ is the maximum amount of RoVs collected
\STATE Insert  $\{[A_0, B_0]$, $[A_1, B_1]$, $...$, $[A_z, B_z]\}$ as TBCs in hardware design
\STATE Test  $\{[A_0, B_0]$, $[A_1, B_1]$, $...$, $[A_z, B_z]\}$
\STATE Insert $LFSR$ algorithm in hardware design
\FOR {image ($X_i$) $\in$ FMV-Dataset}
\STATE $selector = LFSR(X_i[indx])$
\IF {$\{O_i,...,O_{no}\} \in $  $\{[A_0, B_0]$, $[A_1, B_1]$, $...$, $[A_z, B_z]\}$}
\STATE $trig += 1$
\ENDIF
\ENDFOR
\IF {$trig <= threshold$}
\STATE Insert $\{[A_0, B_0]$, $[A_1, B_1]$, $...$, $[A_z, B_z]\}$ as TBC
\ELSE
\STATE Repeat steps 3 to 18
\ENDIF\\
\IF {TBC $\in$ DC as LW trigger}
\STATE Insert TBC as LW trigger
\ELSIF {TBC $\in$ DC as CW trigger}
\STATE Insert TBC as CW trigger
\ELSE
\STATE Insert index-wise trigger
\ENDIF\\
\end{algorithmic}
\label{alg:MYALG1_dynamic}
\end{algorithm}

\begin{figure}[H]
    \centering
    \includegraphics[width=1\linewidth]{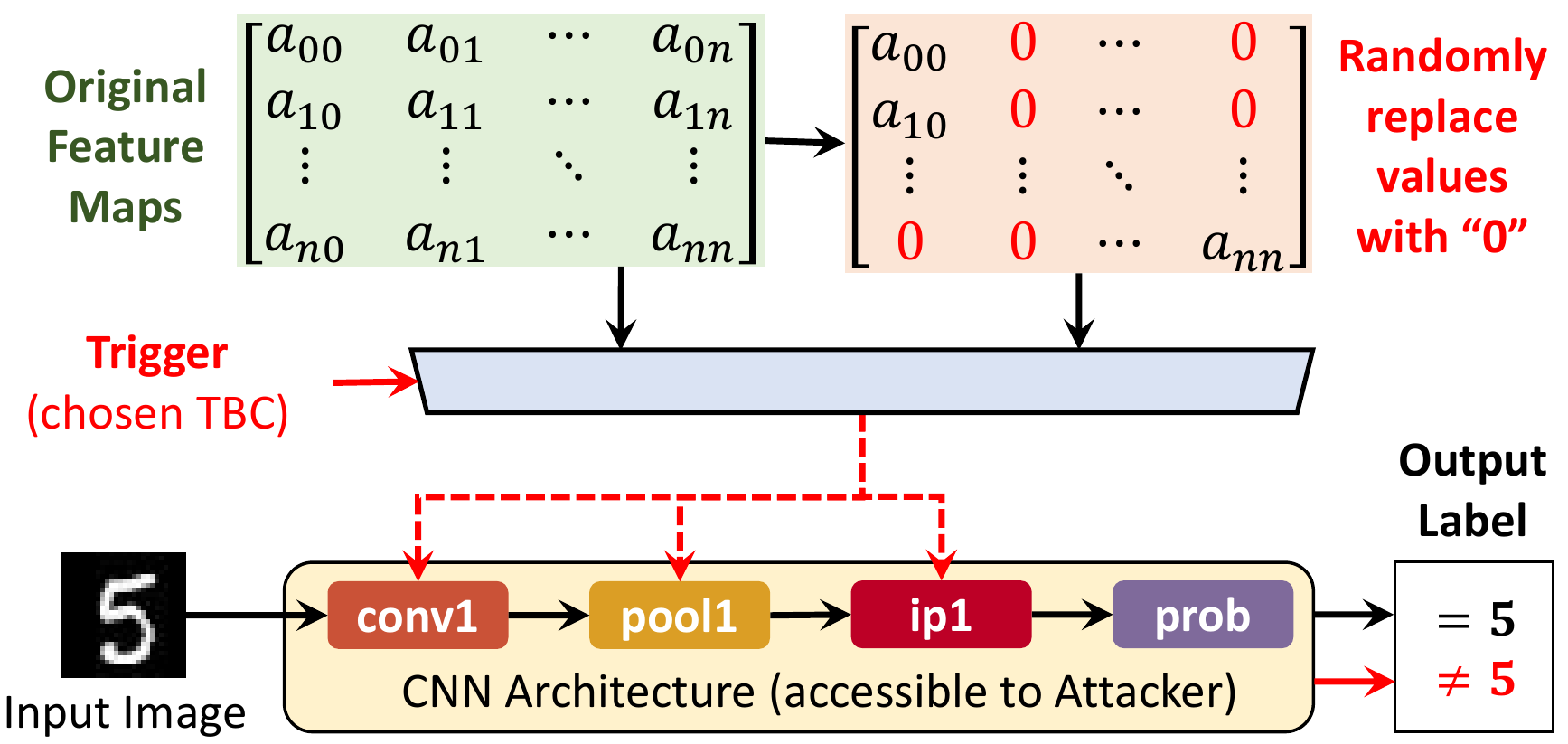}
    \caption{Proposed FeSHI methodology runtime overview: The runtime operation of the hardware Trojan showing the payload. The conversion of feature maps to achieve misclassification}
    \label{overview_run}
\end{figure}

\section{Runtime Operation: Design of the Payload of the Hardware Attack}
\label{payload}
\begin{table*}[t]
    \centering
    \caption{Payload's Empirical Analysis to find the Relationship between the Amount of values rounded down to zero for a given feature map and Loss of Accuracy. This is carried out on LeNet CNN using Vivado HLS 2018.3 for PYNQ FPGA Board (ZYNQ 7020).}
    \label{conv1_conv2}
    \resizebox{1\linewidth}{!}{
    \begin{tabular}{|c|c|c|c|c|c|c|c|c|c|c|c|c|c|c|}
    \hline
    \multirow{3}{*}{Layer} & \multirow{3}{*}{Feature Map} & \multirow{3}{*}{No.} & \multicolumn{12}{c|}{Accuracy loss}                                                      \\ \cline{4-15} 
                           &                        &                           & \multicolumn{4}{c|}{10\%} & \multicolumn{4}{c|}{50\%} & \multicolumn{4}{c|}{90\%} \\ \cline{4-15} 
                           &                        &                           & \#   & FF  & LUT  & Lat.  & \#   & FF & LUT & Lat.  & \#   & FF  & LUT & Lat.  \\ \hline
    $conv1$                &   6 x 24 x 24                     &        $3456$                   &   $1008$   &  $35370$   &  $40279$    &   $610688$    &   $2304$   &  $35766$   &  $40512$    &  $610796$      &  $3168$    &  $35761$   &  $40536$    &   $610868$    \\ \hline
    $conv2$                &     16 x 8 x 8                  &         $1024$                  &   $384$   & $35740$    &  $40776$    &   $601046$    &   $640$   &  $35742$   &  $40779$    &   $601058$    &   $896$   &  $35748$   &   $40802$   &   $601066$    \\ \hline
    
    \end{tabular}}
     \begin{tablenotes}
        \item \textbf{Note,} \textit{No.} represent the total number of values in feature maps, \# represents the total number of values rounded  to zero and \textit{Lat.} represent latency in terms of clock-cycles.
    \end{tablenotes}

\end{table*}

\begin{algorithm}[]
    \footnotesize
    \captionsetup{font={sf,footnotesize},labelfont={bf,color=accessblue}}
    \caption{Payload Operation}
    \begin{algorithmic} [1]
        \REQUIRE CNN Deployment
        \FOR {each image cycle ($X_i$)}
        \IF {$(\{O_i,...,O_z\} \in [a, b])$ (when triggered)}
        \FOR {$i \rightarrow y$ (where $y < z$)}
        \STATE $O_i = 0$
        \ENDFOR
        \ENDIF
        \ENDFOR\\
        $y$ represent number of values in the feature maps rounded down to zero\\
        $z$ represents total number of values in the feature maps of a layer
    \end{algorithmic}
    \label{alg:MYALG1_runtime}
\end{algorithm}

In this work, the payload of the hardware attack is designed to change the values of a feature map of a targeted CNN layer. The proposed payload randomly rounds a number of elements across all channels of the targeted feature map to zero as seen in Lines $2 - 6$ in Algorithm~\ref{alg:MYALG1_runtime} depending on the amount of accuracy loss desired. This is depicted in Fig.~\ref{overview_run}. Since in our threat model, the attacker may not know the exact location of the CNN layer in the overall CNN architecture where they are conducting the attack. Therefore, we demonstrated an analysis of the example network to understand how difficult it is for the attacker to conduct the attack. To achieve this understanding, an empirical study is carried out on LeNet (for MNIST dataset - LeNet architecture is shown in  Fig.~\ref{fig:CNN_MOdel}) to determine the minimum number of values ($y$ shown in Line 3 of Algorithm~\ref{alg:MYALG1_runtime}) of the output feature map that needs to be rounded to zero across all the channels of the targeted CNN layer to guarantee misclassification at the final output. After this determination, the minimum number of values for the output feature maps of interest is inserted into the hardware design. The payload can be inserted into one or more CNN layer(s). Table~\ref{conv1_conv2} shows results of the empirical study of the number of values in a feature map of a layer to be rounded to zero to cause a decrease in accuracy of the mapped CNN (10\%, 50\%, 90\% accuracy decrease respectively) and their corresponding resource utilization for $conv1$ and $conv2$ of LeNet CNN. This empirical study is carried out on only two layers because our goal is just to validate the effect of the proposed payload on high resource requirement layers (in this case, $conv1$ and $conv2$ layers). For the $conv1$ layer, to achieve $10\%$, $50\%$ or $90\%$ accuracy loss, $29\%$, $67\%$ or $92\%$ of the values of the feature maps required to be rounded down to $zero$ respectively. 

For the $conv2$ layer, to achieve $10\%$, $50\%$ or $90\%$ accuracy loss, $38\%$, $63\%$ or $88\%$ of the values of the feature maps required to be rounded down to $zero$ respectively. The result shows that to achieve high accuracy loss, more feature map locations are needed to be rounded to $zero$. The hardware resources required to achieve $90\%$ accuracy loss for both $conv1$ and $conv2$ layers show a maximum percentage difference of 14\% from the original implementation using this payload.

\begin{figure}[]
\centering
\includegraphics[width=1\linewidth]{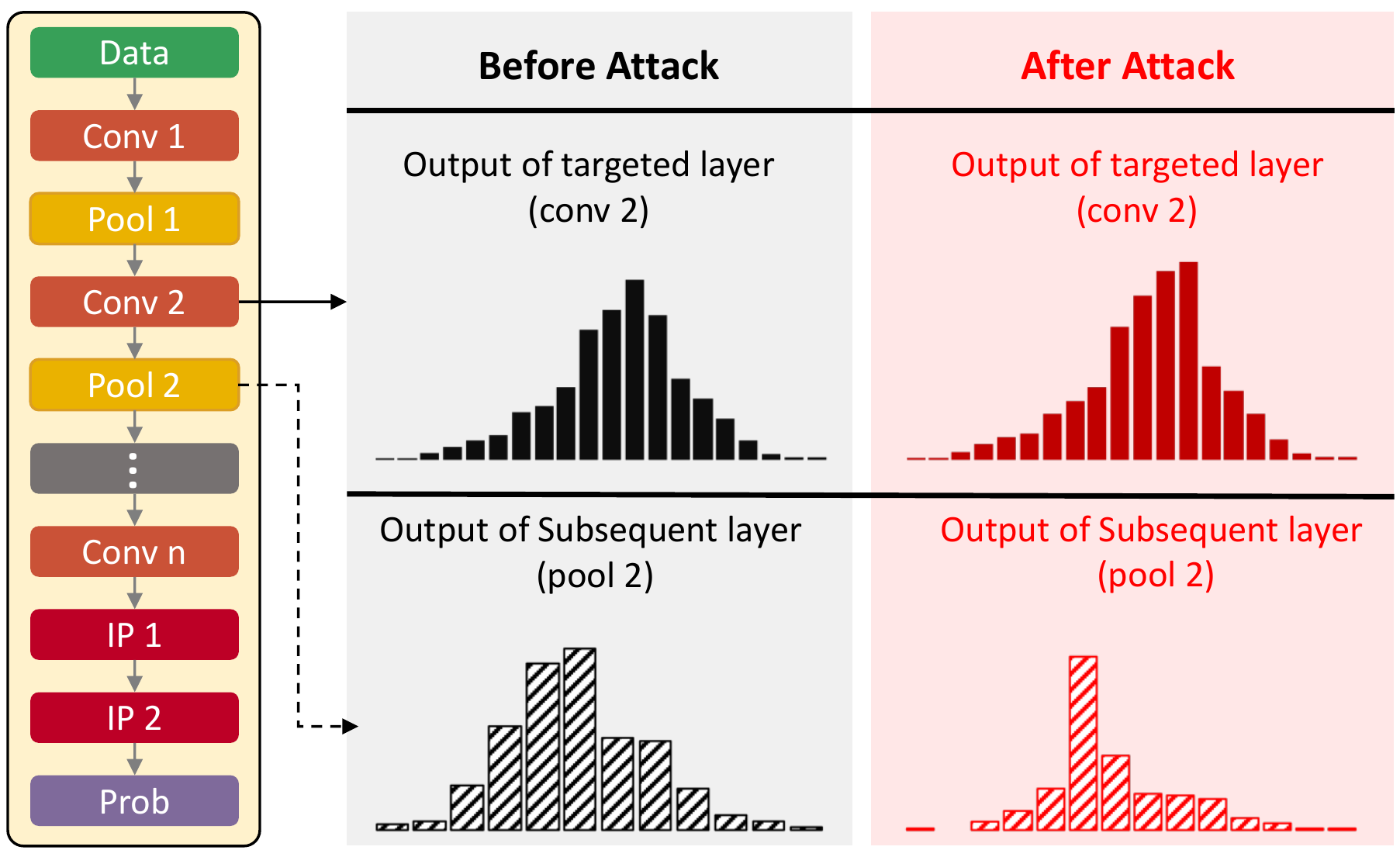}
\caption{Proposed FeSHI methodology runtime overview: runtime operation of the hardware Trojan showing the payload. }
\label{pool_analysis}
\end{figure}

In Fig.~\ref{pool_analysis}, the targeted layer is $conv 2$. When the attack is not triggered (original design) in the conv 2 layer, the statistical distribution of the output feature map of the targeted layer and subsequent layer is shown in Fig.~\ref{pool_analysis}. When the attack is triggered, the proposed payload rounds down $50\%$ of the values of the output feature map causing changes in the statistical distribution in the targeted layer. This change in statistical distribution is cascaded to the next layer (in this case, $pool 2$ layer) and consequently subsequent layers after $pool 2$, which will cause misclassification. 

Table~\ref{conv1_conv2} shows that the number of resources needed (in terms of flip-flops (FF), lookup tables (LUT), and latency (Lat)) is proportional to the number of values needed to be rounded down to zero.  This helps the attacker evaluate the effectiveness of the payload and come up with a stealthiness-effectiveness compromise. 


\begin{table*}[t]
\caption{Hardware resource utilization comparison for both static and dynamic attack on LeNet CNN for MNIST dataset. The BRAM and DSP resource utilization shows $0\%$ percentage difference across all attack scenarios for both static and dynamic attack respectively}
 & 100  & 41562 & 5.52                                             & 36718 & 2.84                                             & 596033                                                         & 0.01                                             & 1.509                                                   & 0.27                                             & 1.655                                                  & 0.42                                             \\ \cline{3-14} 
                                                                                    &                                                                                   & Index                                                    & 1    & 41356 & 5.00                                             & 36653 & 2.66                                             & 595197                                                         & 0.15                                             & 1.513                                                   & \textbf{0.53}                                             & 1.651                                                  & 0.18                                             \\ \hline
\end{tabular}
\begin{tablenotes}
        \item[\#No] \#No. represent the total number of values in feature maps 
    \end{tablenotes}
\label{model_lenet}
\end{table*}

\begin{table*}[t]
\caption{Hardware resource utilization comparison for both static and dynamic attack on LeNet-3D CNN for Cifar10 dataset. The BRAM and DSP resource utilization shows $0\%$ percentage difference across all attack scenarios for both static and dynamic attack except for one scenario of dynamic attack ($conv1-channel$) which shows $0.38$ percentage difference}
 & 100  & 56321                     & 0.00                                                                  & 72818                    & 0.00                                                                  & 1441914                                                                             & 0.00                                                                  & 1.703                                                                        & 0.40                                                                  & 1.865                                                                       & 0.40                                                                  \\ \cline{3-14} 
                                                                                       &                                                                                   & Index                                                    & 1    & 56321                     & 0.00                                                                  & 72818                    & 0.00                                                                  & 1441914                                                                             & 0.00                                                                  & 1.71                                                                         & 0.00                                                                 & 1.872                                                                       & 0.00                                                                  \\ \hline
\end{tabular}
 \begin{tablenotes}
        \item[\#No] \#No. represent the total number of values in feature maps 
    \end{tablenotes}
\label{model_lenet3d}
\end{table*}

\section{Experimental Setup for Hardware Deployment, Results, and Discussion}
\label{results}
\begin{figure}[]
    \centering
    \includegraphics[width=0.75\linewidth]{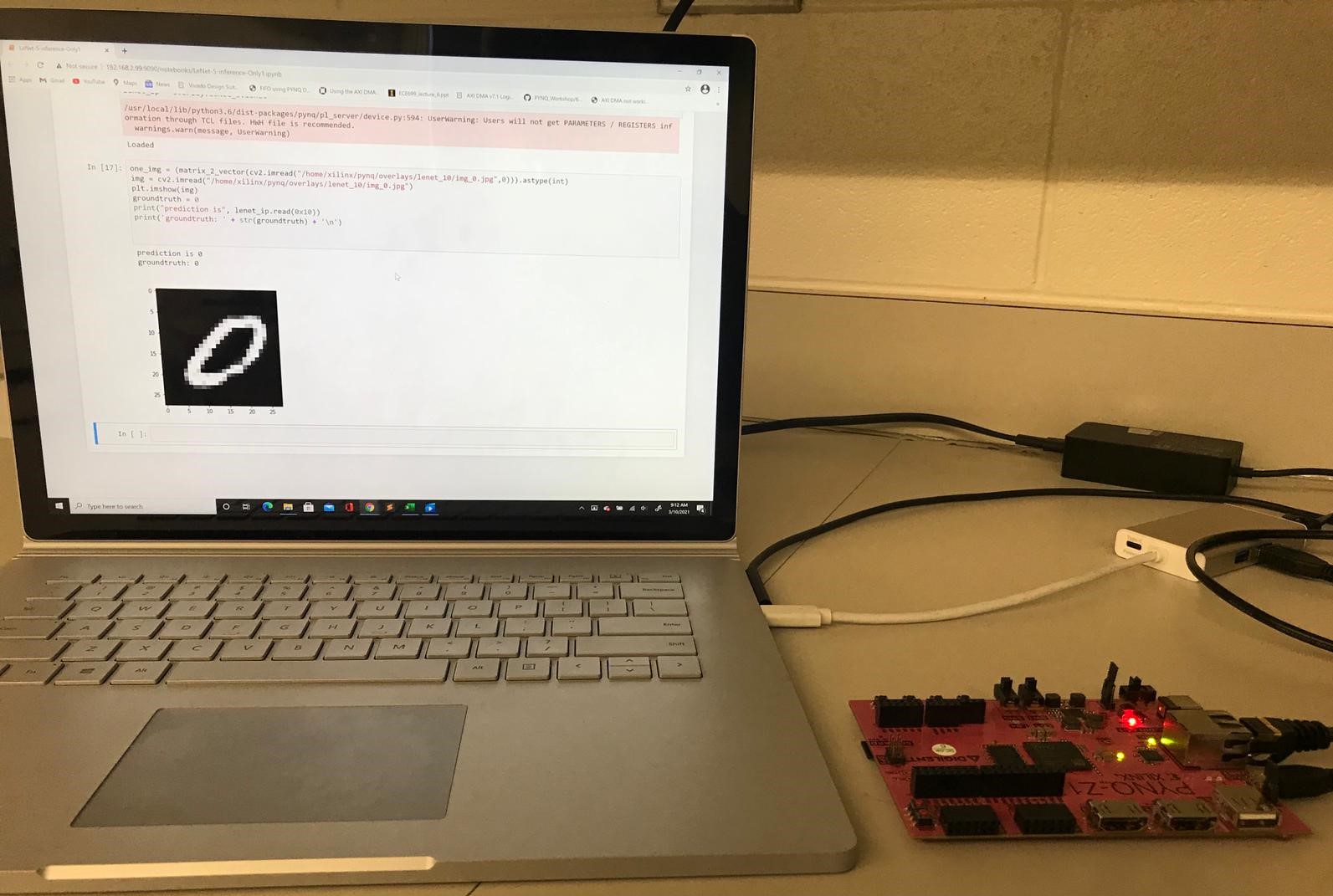}
    \caption{Experiment Setup showing realtime classification of MNIST dataset on PYNQ FPGA using python overlay}
    \label{expt_setup}
\end{figure}

The hardware design of the CNN IP is designed using Xilinx's Vivado and Vivado HLS 2018.3 and to generate an IP for resource-constrained devices. Vivado 2018.3 is used to integrate the generated IP with  AXI-interconnects and  ZYNQ processor (FPGA ZCU 7020 with clock frequency 100MHz) and implemented on the PYNQ FPGA board. The attack is carried out on LeNet CNN trained on MNIST dataset and LeNet-3D CNN trained on Cifar10 datasets as shown in Fig.~\ref{fig:CNN_MOdel}, respectively. The experimental setup is shown in Fig.~\ref{expt_setup}

In this work, we propose $6$ different scenarios (Scenario 1 ($Sn1$) - Scenario 6 ($Sn6$)) of both static  and dynamic attacks, where each layer  (from $conv1$ to $ip1$) is infected with the attack. Table~\ref{model_lenet} shows results of the static and dynamic attack on the LeNet CNN model. For both static and dynamic attacks across all the attack scenarios ($Sn1 - Sn6$) and variations (layer/channel/index), the BRAM and DSP utilization remain the same ($0\%$ difference) when compared to the original implementation. In this work, we have chosen to round down 90\% of the feature map values to zero as a worst-case payload for LeNet and LeNet-3D models with results based on the additional resources and latency overhead shown in Tables~\ref{model_lenet} and~\ref{model_lenet3d}, respectively. \textbf{From  Table~\ref{model_lenet} for the static attack,} LUT utilization across $Sn1 - Sn6$ show a maximum of $6.83\%$ percentage difference compared to the original, which is in the case of $Sn3$ channel attack. FFs across $Sn1 - Sn6$ show a maximum of $1.92\%$ percentage difference compared to the original, which is in the case of $Sn1$ channel attack and $Sn3$ layer attack. Across $Sn1 - Sn6$, the latency shows a maximum percentage difference of $7.15\%$ compared to the original, which is in the case of $Sn1$ layer attack. Both the dynamic and total power usage shows a maximum of $0.53\%$ and $0.49\%$ percentage differences, respectively, when compared to the original. \textbf{In the case of the dynamic attack}, across all the attack scenarios, $Sn1 - Sn6$, LUT utilization shows a maximum of $5.92\%$ percentage difference, the FFs usage shows a maximum of $2.94\%$, the latency shows the highest percentage difference of $7.08\%$, the dynamic power shows a  maximum percentage difference of $0.53\%$ and the total power usage shows a maximum percentage difference of $0.42\%$  when respectively compared to the original (shown in the first row of Table~\ref{model_lenet}). Hence, from these attack scenarios, FeSHI static and dynamic attacks across the LeNet CNN layers are stealthy due to the negligible amount of additional hardware resources required to carry out the attack.

\textbf{Table~\ref{model_lenet3d} shows results of the static and dynamic attacks on the LeNet-3D CNN model.} The result shows that BRAM and DSP usage remains the same, both static and dynamic attacks across all the attack scenarios and stages except for one scenario of dynamic attack ($conv1 channel$), which shows $0.38\%$ percentage difference when compared to the original. Also, from  Table~\ref{model_lenet3d}, for the static attack, LUT utilization across $Sn1 - Sn6$ shows a maximum percentage difference of $4.48\%$ compared to the original, which is in the case of $Sn3$ channel attack. FFs across $Sn1 - Sn6$ show a maximum percentage difference of $1.10\%$ when compared to the original, which is in the case of the $Sn2$ index attack. Across $Sn1 - Sn6$, the latency shows a maximum percentage difference of $6.86\%$ compared to the original, which is in the case of $Sn1$ layer attack. Dynamic power usage shows a maximum percentage difference of $2.63\%$ when compared to the original, which occurs in the case of a $Sn1$ channel attack. While total power usage shows a maximum percentage difference of $2.51\%$ when compared to the original, which occurs in the case of $Sn1$ layer and channel attack.

\textbf{In the case of the dynamic attack for LeNet-3D,} across all the attack scenarios, $Sn1 - Sn6$, LUT utilization shows a maximum of $3.17\%$ percentage difference in $Sn1$ channel attack, and the FFs usage shows a maximum of $0.21\%$ in $Sn1$ channel attack and $Sn1$ layer and channel attack, the latency shows the highest percentage difference of $3.55\%$ in the case of $Sn1$ layer attack, the dynamic power shows a percentage difference of maximum of $2.57\%$ in $Sn1$ channel attack and the total power usage shows a maximum of $2.72\%$ percentage difference in $Sn1$ channel attack when compared respectively compared to the original. These results confirm that the proposed FeSHI static and dynamic attacks across the LeNet-3D CNN layers are stealthy due to the negligible additional hardware resources required to carry out the attack.

\begin{figure}[!t]
    \begin{subfigure}{\linewidth}
      \centering
      \includegraphics[width=1\linewidth]{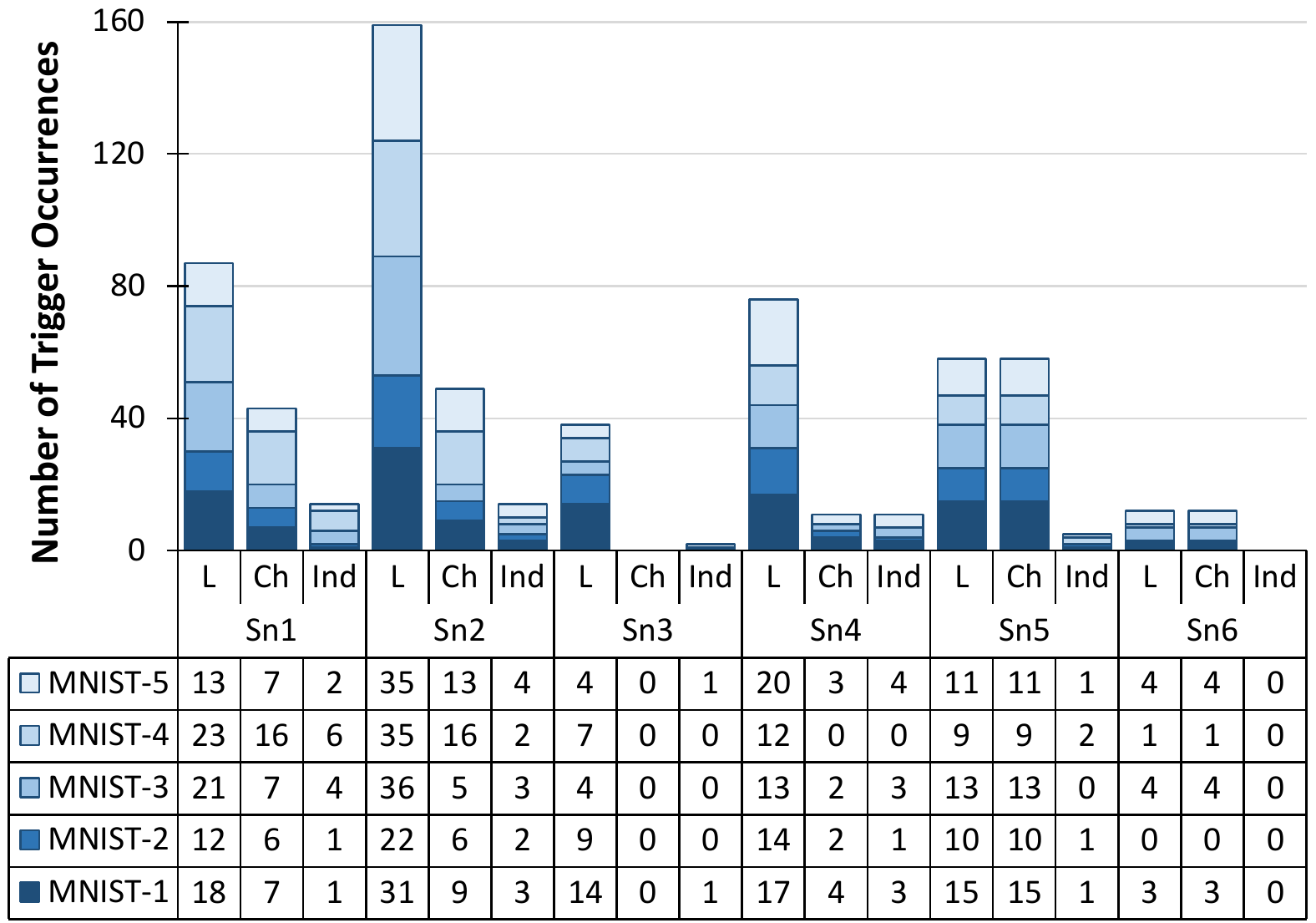}  
      \caption{Different variation of Static FeSHI attack (Layer-wise (L), Channel-wise (Ch) and Index-wise (Ind)) on LeNet under different scenarios.}
      \label{stealth_test1_a}
    \end{subfigure}
    \begin{subfigure}{\linewidth}
      \centering
      \includegraphics[width=1\linewidth]{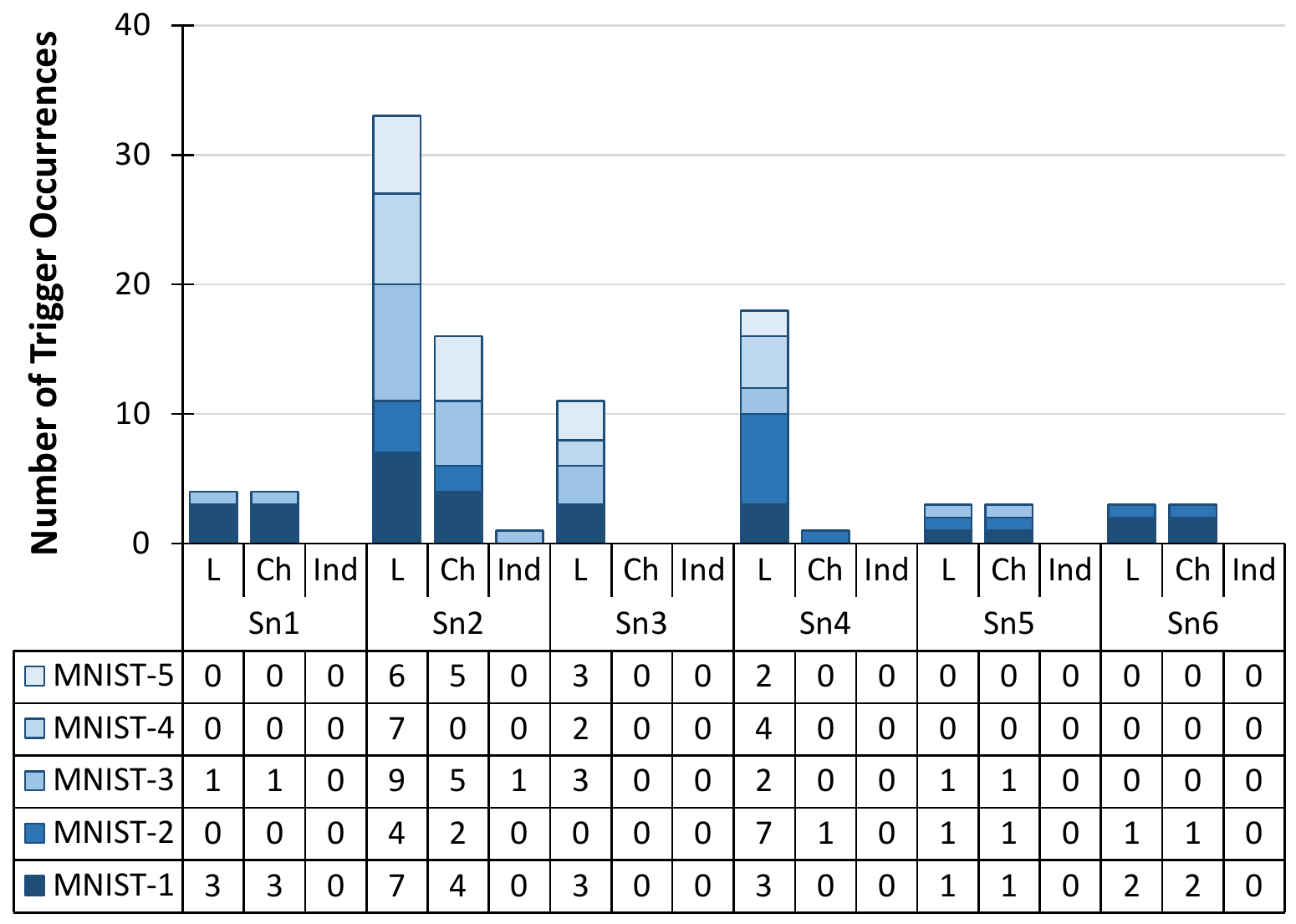}  
      \caption{Different variation of Dynamic FeSHI attack (Layer-wise (L), Channel-wise (Ch) and Index-wise (Ind)) on LeNet under different scenarios.}
      \label{stealth_test1_b}
    \end{subfigure}
    \caption{Experimental results to show the random nature and low-triggering rate of FeSHI attack (Layer-wise (L), Channel-wise (Ch) and Index-wise (Ind)) on \textbf{LeNet trained on MNIST dataset} under different scenarios.}
    \label{stealth_test1}
\end{figure}

\begin{figure}[!t]
    \begin{subfigure}{\linewidth}
      \centering
      \includegraphics[width=1\linewidth]{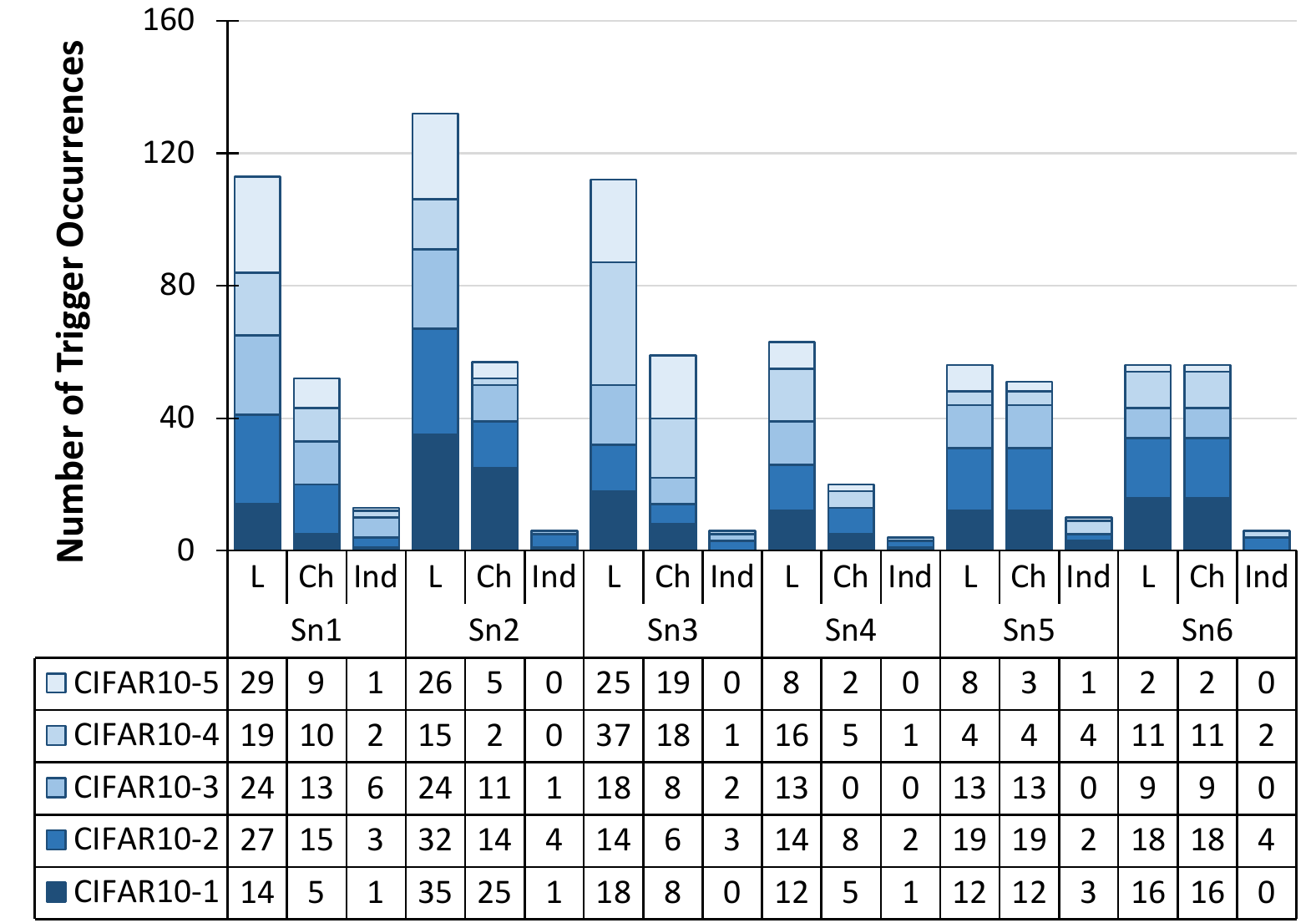}  
      \caption{Different variation of Static FeSHI attack (Layer-wise (L), Channel-wise (Ch) and Index-wise (Ind)) on LeNet-3D under different scenarios.}
      \label{stealth_test2_a}
    \end{subfigure}
    \begin{subfigure}{\linewidth}
      \centering
      \includegraphics[width=1\linewidth]{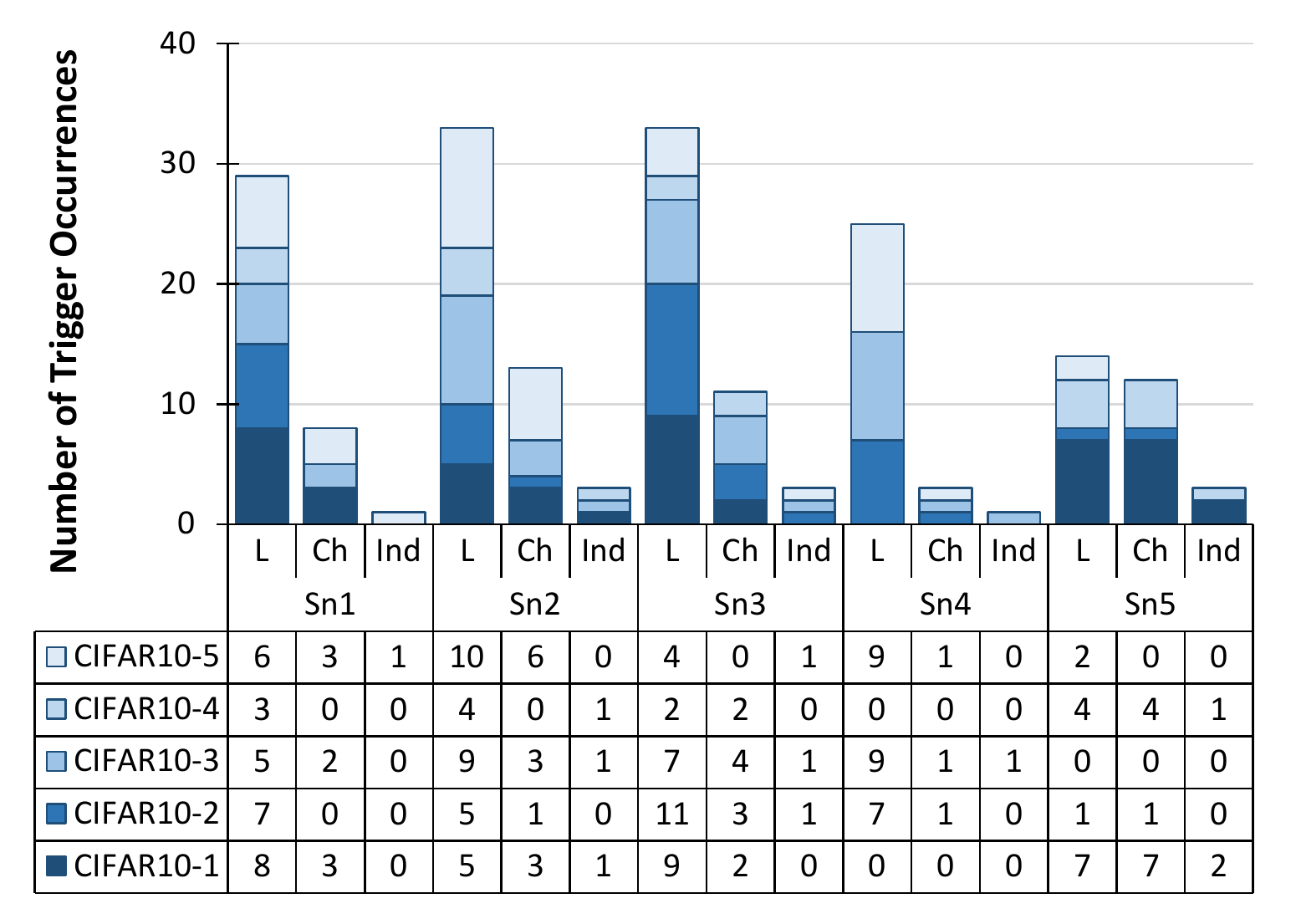}  
      \caption{Different variation of Dynamic FeSHI attack (Layer-wise (L), Channel-wise (Ch) and Index-wise (Ind)) on LeNet-3D under different scenarios.}
      \label{stealth_test2_b}
    \end{subfigure}
    \caption{Experimental results to show the random nature and low-triggering rate of FeSHI attack (Layer-wise (L), Channel-wise (Ch) and Index-wise (Ind)) on \textbf{LeNet-3D trained on CIFAR10 dataset} under different scenarios.}
    \label{stealth_test2}
\end{figure}

To demonstrate the randomness in the triggering of the proposed attack, various random input datasets are examined. In Figs.~\ref{stealth_test1} and~\ref{stealth_test2}, we randomly select and provide five different datasets (200 images each) to LeNet, and LeNet-3D CNNs, the static and dynamic attack variations are investigated from the layer-wise, channel-wise, and index-wise perspectives. For and LeNet CNN, considering $Sn1$, the number of trigger occurrences varies randomly between 10 and 20 in the layer-wise triggering approach, 4 and 15 in the channel-wise triggering approach, 0 and 5 in the index-wise triggering approach. The same is true for other attack scenarios- making our attack random and stealthy. During design time, we experienced certain triggering probability on seen data (FMV-Dataset).

\subsection{Possible Defense Mechanisms}
Although these stealthy attacks can successfully perform misclassification, one possible defense mechanism against the proposed FeSHI static attack is corner case analysis. The defender can perform the corner case analysis during the testing phase to detect such HT-based attacks. In the corner case analysis-based defense, the outlier samples, which have a higher possibility of triggering the attack, are used to test the outsourced hardware modules. Hence, the system integrator can detect the static version of the FeSHI attack, with one constant triggering condition, by carefully designing the testing dataset. To make the attack stealthier against the above-mentioned corner case analysis-based defense,  we proposed a dynamic version of the FeSHI attack. In this attack, the triggering condition changes randomly between 5 preselected possible intervals, which are obtained apriori from the $3 \sigma$ analysis. The dynamic version of the FeSHI attack does not require enormous additional hardware resources and proves to be stealthier compared to the static version of the FeSHI attack.

\textbf{Possible Defense Against Dynamic FeSHI Attack:} The dynamic version of the FeSHI attack is very effective and stealthy. However, if the system integrator is aware of such an attack scenario, then the following uncorrelated FMV-Dataset-based defense approach can be used to mitigate the attack:
In this defense, the defender can exploit the fact that for the correct implementation of hardware design, the correlation between the datasets used for validation (in this case, FMV-Dataset) and the dataset used for CNN training, validation, and inference at the software level is not compulsory. Therefore, the defender can provide completely new and unrelated FMV-datasets such that the correlation between the input and output data instances of the FMV-Datasets and the input and output instances of the training, testing, or validation datasets are approximately zero. Using the uncorrelated FMV-Dataset, the attacker would design the triggering conditions that may never occur during the run-time CNN inference; hence, HT-based attack would possibly be mitigated.

\section{Conclusion and Future Works}
In this paper, we propose a hardware Trojan attack that performs a statistical analysis on the layer-by-layer output of CNN models to obtain stealthy triggering conditions. This attack is a grey box attack that is carried out in the hardware implementation stage of CNN architecture. This attack is demonstrated on two CNN models used to perform classification on MNIST and CiFAR-10 datasets. The Trojan proved to be stealthy and adds negligible hardware overhead.
\bibliographystyle{IEEEtran}
\bibliography{IEEEabrv,bib/Bibliography}
\appendix
\section{}
\label{appendix}
\begin{figure*}[]
    \centering
    \includegraphics[width=1\linewidth]{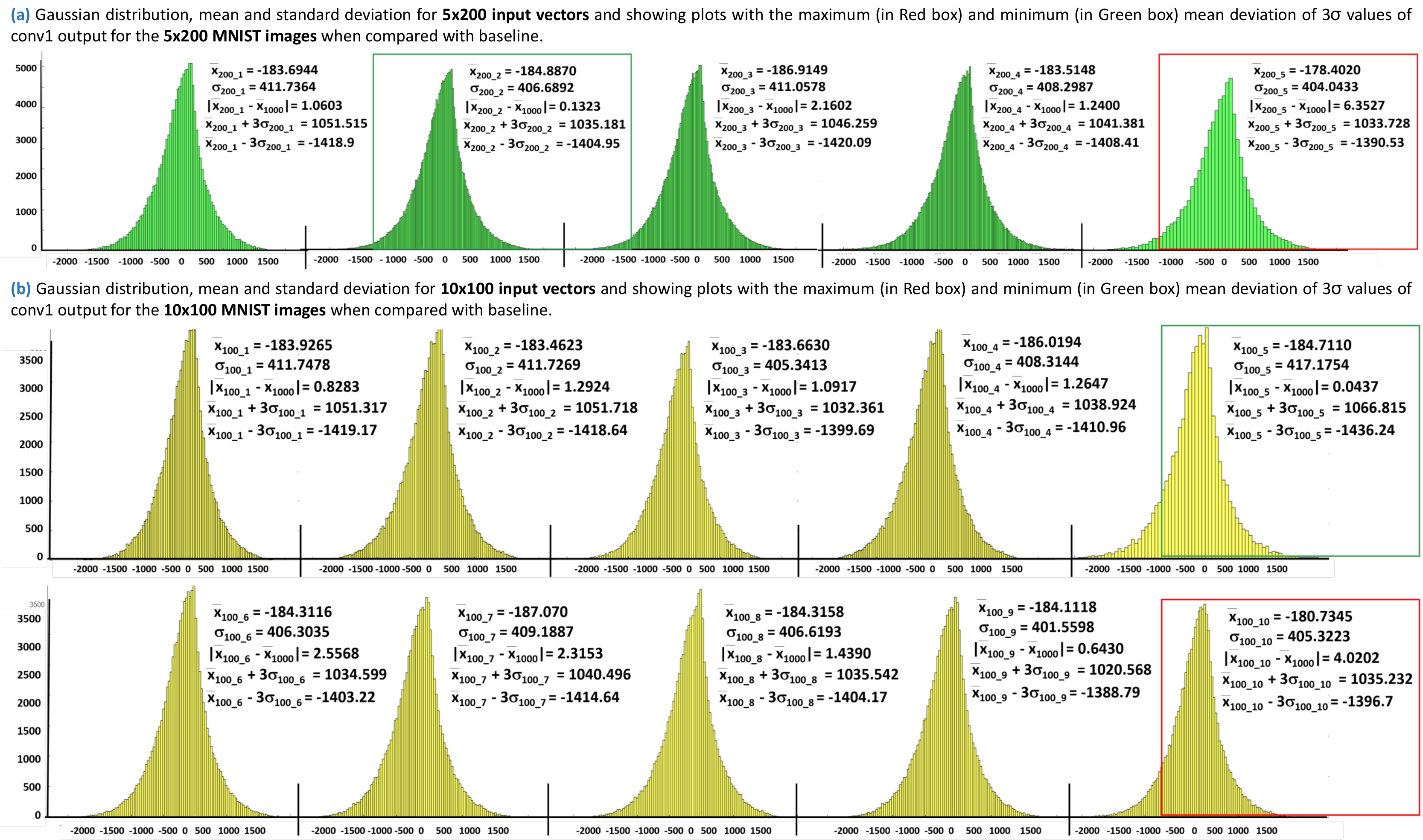}
    \caption{Comparison of mean, standard deviation, Gaussian distribution and $3\sigma$ values of different sizes of datasets input images for LeNet CNN trained for MNIST dataset. These plots validate that with reduced dataset from 1000 input vectors to 10x100 and 5x200 input vectors. The minimum and maximum mean deviations also fall around the same range.}
    \label{input_vectors_appendix1}
\end{figure*}

\begin{figure*}[]
    \centering
    \includegraphics[width=1\linewidth]{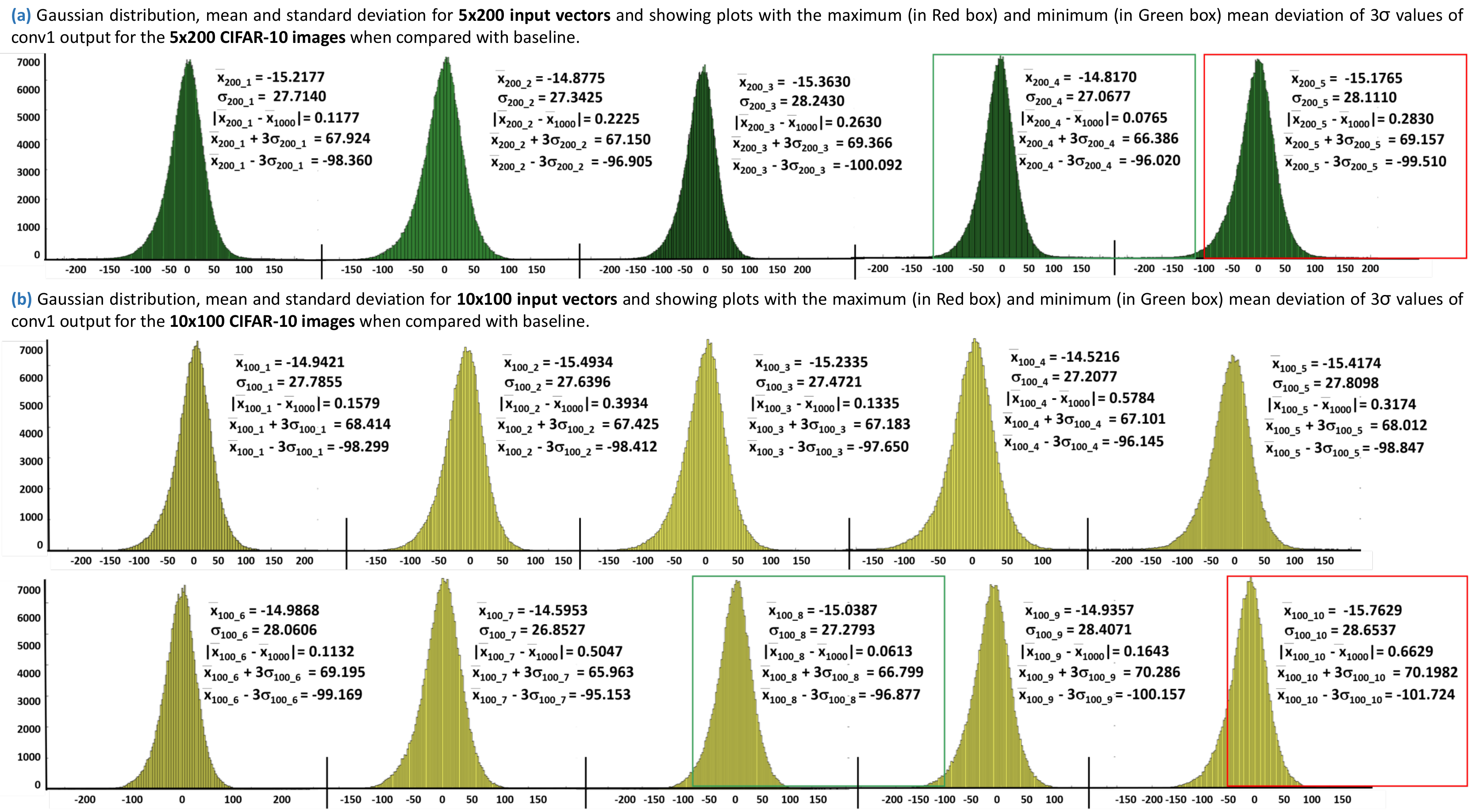}
    \caption{Comparison of mean, standard deviation, Gaussian distribution and $3\sigma$ values of different sizes of datasets input images for LeNet-3D CNN trained for Cifar-10 dataset. These plots validate the hypothesis that with reduced dataset from 1000 input vectors to 10x100 and 5x200 input vectors. The minimum and maximum mean deviations also fall around the same range.}
    \label{input_vectors_appendix2}
\end{figure*}

From the detailed statistical analysis of FMV-Dataset (Section~\ref{size of dataset}), we observed the following:
\begin{itemize}[leftmargin=*]
    \item The Gaussian distribution, mean and standard deviation  of and 5$\times$200 and 10$\times$ 100 input vectors respectively of MNIST images (as seen in Fig. \ref{input_vectors_appendix1}. Within the subplots, the distribution with the  maximum (in Red box) and minimum (in Green box) mean deviation of $3\sigma$ values when compare to the baseline in Fig. \ref{input_vectors1}. It can be seen that even the maximum $3\sigma$ deviation, which helps this technique in defining RoVs (see Section \ref{trigger}), is within $1.40\%$ and $2.50\%$ of the baseline values, respectively. 
    
    \item The Gaussian distribution, mean and standard deviation  of and 5$\times$200 and 10$\times$ 100 input vectors respectively of Cifar10 images (as seen in Fig. \ref{input_vectors_appendix2}. Within the subplots, the distribution with the  maximum (in Red box) and minimum (in Green box) mean deviation of $3\sigma$ values when compare to the baseline in Fig. \ref{input_vectors2}. It can be seen that even the maximum $3\sigma$ deviation, which helps this technique in defining RoVs (see Section \ref{trigger}), is within $3.59\%$ of the baseline values.
\end{itemize}


\begin{IEEEbiography}[{\includegraphics[width=1in,height=1.25in,clip,keepaspectratio]{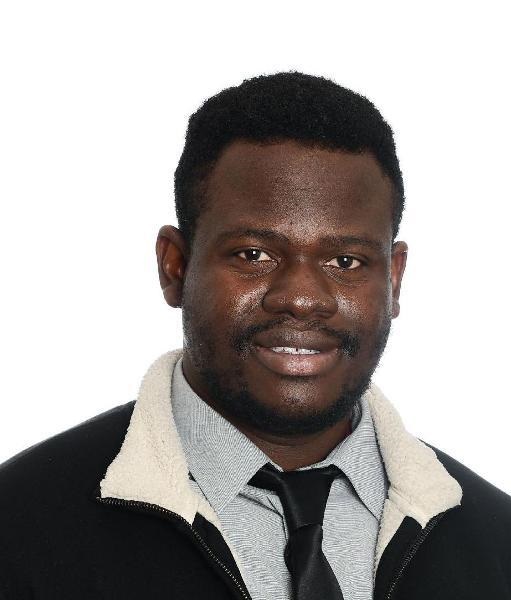}}]
{Tolulope A. Odetola} received his Bachelors degree from Obafemi Awolowo University Nigeria in Electronic and Electrical Engineering. He received his Masters degree in Electrical and Computer Engineering from Tennessee Technological University, Cookeville, TN, USA. He is currently pursuing his Doctorate degree in Tennessee Technological University. He developed hardware/software co-verification techniques for CNN deployments on hardware accelerators.  His current research interest includes FPGAs, hardware security, IoT security, machine learning, and deep learning.
\end{IEEEbiography}
\begin{IEEEbiography}[{\includegraphics[width=1in,height=1.25in,clip,keepaspectratio]{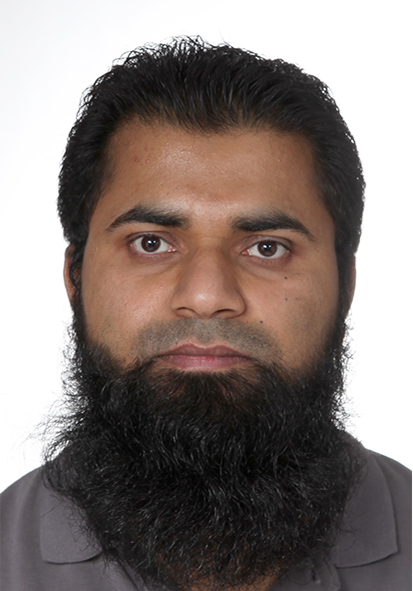}}]{Faiq Khalid}(GS'18-M'21) received his M.S. degree in electrical engineering and his B.E. degree in electronics engineering from the National University of Sciences and Technology (NUST), Pakistan, in 2016 and in 2011, respectively. He is currently pursuing his Ph.D. degree in hardware security and machine learning security at Technische Universit{\"a}t Wien (TU Wien), Vienna, Austria. He is a recipient of the Quaid-e-Azam Gold Medal for his academic achievements and the Richard Newton Young Fellowship Award at DAC 2018. His research interests include formal verification of embedded systems, hardware design security, and security for machine learning systems. He has also served as a TPC member of FIT, WSAV, ARES and ICONS.   
\end{IEEEbiography}
\begin{IEEEbiography}[{\includegraphics[width=1in,height=1.25in,clip,keepaspectratio]{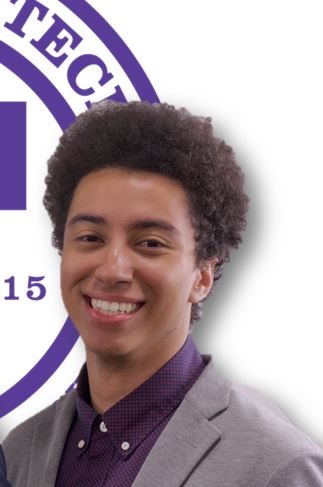}}]
{Travis Sandefur} received his Bachelors in Electrical and Computer Engineering from Tennessee Technological University, Cookeville, TN, USA. He is currently pursuing his masters in Electrical and Computer Engineering also in Tennessee Technological University. His currently working on techniques for object detection on resource constrained devices.  His current research interest includes FPGAs, machine learning, partial reconfiguration and deep learning.
\end{IEEEbiography}
\begin{IEEEbiography}[{\includegraphics[width=1in,height=1.25in,clip,keepaspectratio]{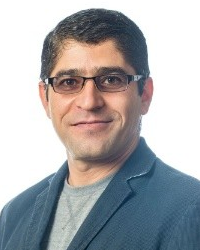}}]
{Hawzhin Mohammed} received his B.Sc. degree in electrical engineering from Salahaddin University, Erbil, Iraq, in 2000. He received his M.Sc. degree from Tennessee Technological University, Cookeville, TN, USA, in 2017, where he is currently pursuing his Ph.D. degree at the Department of Electrical and Computer Engineering. His current research interest includes wireless network security, hardware security, IoT security, machine learning, and deep learning.
\end{IEEEbiography}
\begin{IEEEbiography}[{\includegraphics[width=1in,height=1.25in,clip,keepaspectratio]{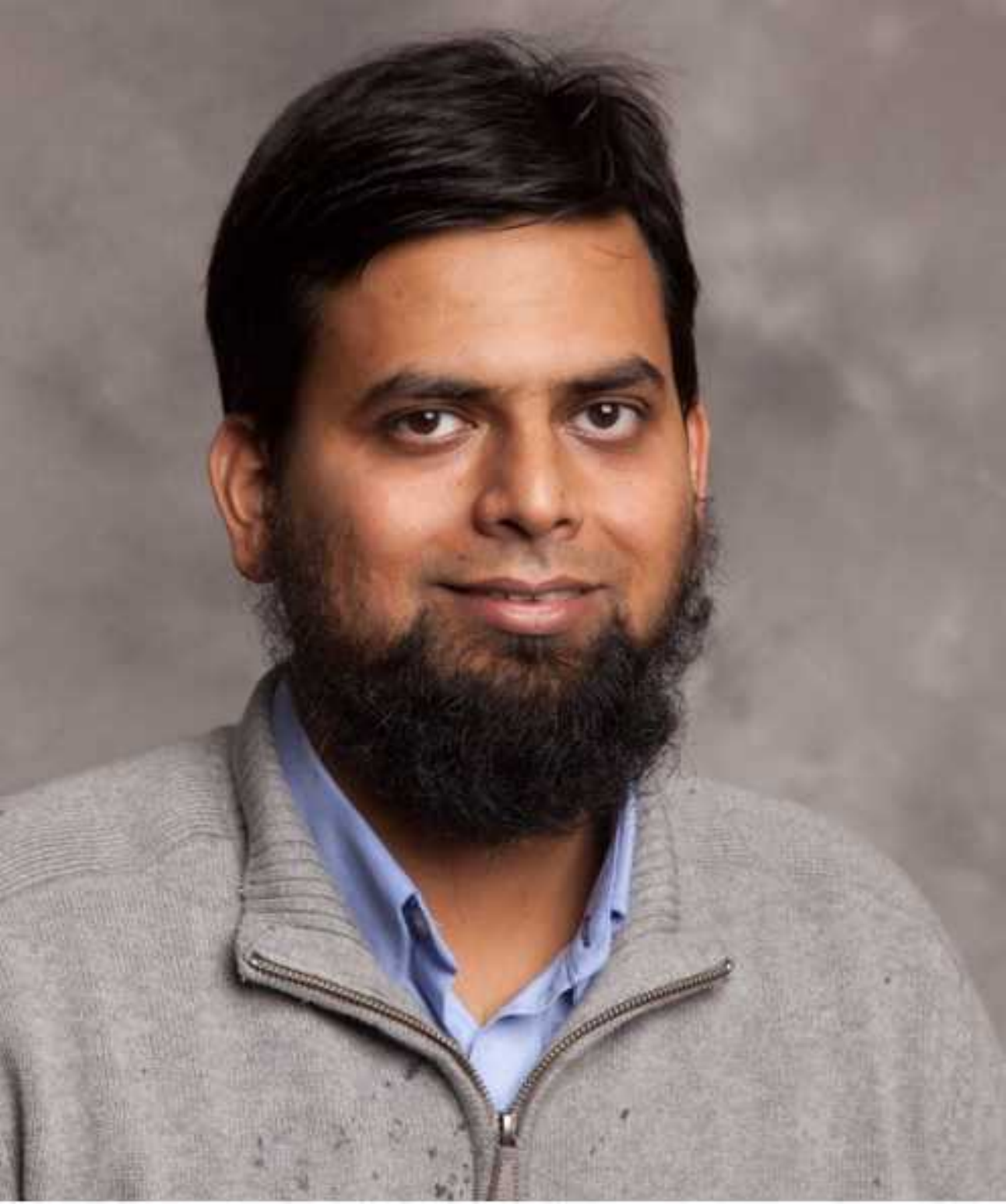}}]
{Syed Rafay Hasan} received the B.Eng. degree in electrical engineering from the NED University of Engineering and Technology, Pakistan, and the M.Eng. and Ph.D. degrees in electrical engineering from Concordia University, Montreal, QC, Canada. From 2006 to 2009, he was an Adjunct Faculty Member with Concordia University. From 2009 to 2011, he was a Research Associate with the Ecole Polytechnique de Montreal. Since 2011, he has been with the Electrical and Computer Engineering Department, Tennessee Tech University, Cookeville, TN, USA, where he is currently an Associate Professor. He has published more than 69 peer-reviewed journal and conference papers. His current research interests include hardware design security in the Internet of Things (IoT), hardware implementation of deep learning, deployment of convolution neural networks in the IoT edge devices, and hardware security issues due to adversarial learning. He received SigmaXi Outstanding Research Award, Faculty Research Award from Tennessee Tech University, the Kinslow Outstanding Research Paper Award from the College of Engineering, Tennessee Tech University, and the Summer Faculty Fellowship Award from the Air force Research Lab (AFRL). He has received research and teaching funding from NSF, ICT-funds UAE, AFRL, and Intel Inc. He has been part of the funded research projects, as a PI or a Co-PI, that worth more than \$1.1 million. He has been the Session Chair and Technical Program Committee Member of several IEEE conferences including ISCAS, ICCD, MWSCAS, and NEWCAS, and a Regular Reviewer for several IEEE Transactions and other journals including TCAS-II, IEEE ACCESS, Integration, the VLSI Journal, IET Circuit Devices and Systems, and IEEE Embedded System Letters.
\end{IEEEbiography}

\EOD
\end{document}